\newif\ifconfver
\confverfalse      %declaring conference version false
\newif\ifcutshort      %this level shortens the equations
\cutshorttrue

\newif\ifcutshortlvltwo  %this level takes out some examples, figs., and sim.
\cutshortlvltwofalse

\ifconfver
    \documentclass[10pt,twocolumn,twoside]{IEEEtran}
\else
    \documentclass[11pt,draftcls,onecolumn]{IEEEtran}
\fi

\usepackage{cite}
\usepackage{graphicx}
\usepackage{psfrag}
\usepackage{subfigure}
\usepackage{url}
\usepackage{stfloats}
\usepackage{amsfonts,amssymb,amsmath,bm,paralist,theorem,cite,ifthen,color}
\usepackage{array}
\usepackage{caption}
\usepackage{calc}
\usepackage{subfig}
\usepackage{longtable}
\usepackage{stfloats}  % Written by Sigitas Tolusis
\usepackage{algpseudocode}
\usepackage{algorithmicx}

\usepackage{graphics,booktabs,color,epsfig,subfigure}

\newtheorem{lemma}{Lemma}
\newtheorem{proposition}{Proposition}

\newtheorem{Theorem}{Theorem}

\newtheorem{Rmk}{Remark}

\theorembodyfont{\rmfamily}

{\begin{list}{}{
    \settowidth{\labelwidth}{\mbox{\textnormal{#1}}}%
    \setlength{\leftmargin}{\labelwidth+\labelsep}}}%
{\end{list}}

\newcommand{\bfmu}{{\mbox{\boldmath $\mu$}}}
\newcommand{\bflambda}{{\mbox{\boldmath $\lambda$}}}
\newcommand\xb{\ensuremath{{\bm x}}}

\newcommand\wb{\ensuremath{{\bm w}}}

\newcommand\ub{\ensuremath{{\bf u}}}
\newcommand\Hb{\ensuremath{{\bf H}}}
\newcommand\hb{\ensuremath{{\bf h}}}
\newcommand\Ab{\ensuremath{{\bf A}}}

\newcommand\Bb{\ensuremath{{\bf B}}}
\newcommand\bb{\ensuremath{{\bf b}}}

\newcommand\cb{\ensuremath{{\bf c}}}

\newcommand\Kb{\ensuremath{{\bf K}}}
\newcommand\Ib{\ensuremath{{\bf I}}}

\newcommand\vb{\ensuremath{{\bf v}}}

\newcommand\zb{\ensuremath{{\bf z}}}

\newcommand\Cplx{\ensuremath{{\mathbb{C}}}}
\newcommand\Ral{\ensuremath{{\mathbb{R}}}}
\newcommand{\cI}{\mathcal{I}}
\newcommand{\cQ}{\mathcal{Q}}

\newcommand{\bfkappa}{{\mbox{\boldmath $\kappa$}}}
\newcommand{\bfalpha}{{\mbox{\boldmath $\alpha$}}}

\begin{document}

\bibliographystyle{IEEEtran}
% paper title

\title{Base Station Activation and Linear Transceiver Design for Optimal Resource Management in Heterogeneous Networks}

\ifconfver \else {\linespread{1.1} \rm \fi

\author{
%\vspace{0.5cm}
Wei-Cheng Liao$^{\dag\star}$, Mingyi Hong$^\dag$, Ya-Feng Liu$^\ddag$, and
Zhi-Quan Luo$^\dag$
\thanks{$^\S$This work is supported in part by the National Science Foundation,
grant number CCF-121685, and by a research gift from Huawei
Technologies Inc. This paper was presented in part at the IEEE
ICASSP 2013\cite{Liao13}.}
\thanks{$^\dag$Wei-Cheng Liao, Mingyi Hong, and Zhi-Quan Luo are with
the Department of Electrical and Computer Engineering, University of
Minnesota, Minneapolis, MN 55455. E-mail:
$\{$liaox146,mhong,luozq$\}$@umn.edu}
\thanks{$^\ddag$Ya-Feng Liu is with the State Key Laboratory of Scientific
and Engineering Computing, Institute of Computational Mathematics and Scientific/Engineering Computing, Academy of Mathematics and Systems Science, Chinese Academy of Sciences, Beijing, 100190, China. E-mail: yafliu@lsec.cc.ac.cn.}
\thanks{$^\star$ Correspondence author.}}

\maketitle

\vspace*{-1.5\baselineskip}

\begin{abstract}
In a densely deployed heterogeneous network (HetNet), the number of
pico/micro base stations (BS) can be comparable with the number of
the users. To reduce the operational overhead of the HetNet, proper
identification of the set of serving BSs becomes an important design
issue. In this work, we show that by jointly optimizing the
transceivers and determining the active set of BSs, high system
resource utilization can be achieved with only a small number of
BSs. In particular, we provide formulations and efficient algorithms
for such joint optimization problem, under the following two common
design criteria: {\it i)} minimization of the total power
consumption at the BSs, and {\it ii)} maximization of the system
spectrum efficiency. In both cases, we introduce a nonsmooth regularizer
to facilitate the activation of the most appropriate BSs.
%We develop efficient algorithms to solve the resulting regularized nonsmooth optimization problems
%by a novel application of the Alternating Direction Method of Multipliers (ADMM) and
%the weighted MMSE reformulation.
We illustrate the efficiency and the efficacy of the
proposed algorithms via extensive numerical simulations.

%For each of the design criteria, the is formulated as a nonsmooth
%(group)-LASSO regularized optimization problem.
%
%Specifically, for the total power minimization design criterion, we
%first show that the considered active BS selection problem is
%NP-hard. Then based on Alternating Direction Method of Multipliers
%(ADMM) algorithm, an efficient algorithm is proposed, which is shown
%to have closed-form update rule in each iteration.
%
%For the system spectrum efficient design criterion, inspired by
%weighted minimum mean square error (WMMSE) algorithm, we propose an
%algorithm that is guaranteed to converge to a stationary solution.
%

\end{abstract}
\begin{keywords}
Heterogeneous networks, LASSO, Base station selection/clustering,
Alternating Direction Method of Multipliers (ADMM), Weighted MMSE algorithm
\end{keywords}
\section{Introduction}
\label{sec:intro}

To cope with the explosive growth of mobile wireless data, service providers
have increasingly relied on adding base stations (BSs) to provide better cell coverage and higher
level of quality of service, resulting in a heterogeneous network (HetNet) architecture. Moreover, the recent LTE-A
standard has also advocated this type of architecture whereby macro BSs are used to cover
large areas, while low-power transmit nodes such as pico/micro BSs are
densely deployed for coverage extension \cite{Damnjanovic11}. The
main strength of this new type of cellular network lies in its
ability to bring the transmitters and receivers close to
each other, so that significantly less transmit power is needed to
deliver higher signal quality and system performance.

%As a result, the architecture of traditional cellular
%networks will undergo a fundamental shift in which a macro base
%station (BS) and various pico/micro BSs may coexist in a cell and
%collaborate to serve the users while sharing the same spectrum

One central issue arising in the HetNet is interference
management, a topic which has attracted extensive research efforts
lately \cite{hong12survey}. Among many existing schemes, node
cooperation appears quite promising. In LTE-A \cite{Clerckx09},
two main modes of cooperation have been considered \cite{Gesbert10}:
{\it (1)} Joint Processing (JP), in which several BSs jointly
transmit to users by sharing transmitted data via high speed
backhaul network; {\it (2)} Coordinated Beamforming (CB), in which
BSs mitigate interference by cooperative transmit
beamformming without sharing users' data. These two
approaches complement each other---JP achieves high spectrum
efficiency, while the CB requires less backhaul capacity.  Recently
there have been many works that propose to strike a balance between
these two approaches, especially in cases where the number of BSs
is large
\cite{Zhang09,Papadogiannis08,Papadogiannis10,Zeng10,Kim12,Hong12,Cheng13}.
The idea is to cluster a small number of BSs together such that JP
is used only within each BS cluster. Although these schemes have
satisfactorily addressed the tradeoff between the effectiveness of
interference management and the signaling overhead,  most of them have
neglected the fact that when large number of BSs are simultaneously
activated, substantial operational costs are incurred
\cite{Gesbert10,Arnold10}. These costs can take the form of power
consumption, complexity for encoding/decoding, or overhead related
to BS management or information exchanges among the BSs. To keep the
operational cost manageable, it is necessary to appropriately
select a subset of active BSs while shutting down the rest. To the
best of our knowledge, except \cite{Cheng13}, none of the existing works on BS clustering
considers this factor in their formulations; see e.g.,
\cite{Zhang09,Papadogiannis08,Papadogiannis10,Zeng10,Kim12,Hong12}.
As a result, the solutions computed by these algorithms typically
require most BSs in the network to remain active. Moreover, although \cite{Cheng13} takes BS activation into consideration, the formulated mixed-integer optimization cannot be efficiently solved for large-scale HetNet.

In this work, we propose to design optimal downlink transmit
beamforming strategies for a HetNet under the following two criteria: {\bf
C1)} given a prescribed quality of service (QoS), minimize the total
power consumption, and {\bf C2)} given the power constraints on each
BS, maximize the sum rate performance. In contrast to the
existing literature on downlink beamforming, we
impose the additional requirement that these design criteria
are met with a small number of BSs. In our formulation, the latter is achieved by
imposing certain sparsity pattern in the users' beamformers. This idea has
also recently been used in different applications in wireless
communications, e.g., antenna selection in downlink transmit
beamforming \cite{Mehanna12}, joint power and admission control \cite{Liu13}, and the joint precoder design with
dynamical BS clustering \cite{Zeng10,Kim12,Hong12}. However,
none of these works seek to reduce the number of active BSs in the
network.

 From the complexity standpoint, the problems being considered are computationally challenging:
we show that the problem of selecting the minimum number of active BSs that satisfy
a given set of QoS constraints is strongly NP-hard for a multi-input single-output
(MISO) system. This motivates us to design practical signal processing algorithms for the problems {\bf C1)} and
{\bf C2)}. To this end, our contributions are two folds. First,
we generalize the traditional power minimization beamforming design (see
\cite{Wiesel06,Dahrouj10}) by formulating problem {\bf C1)} as a second-order
cone program (SOCP) using a sparsity regularizer. Despite the fact that such SOCP admits
a convex representation, direct optimization using standard packages not only
requires central control and a large communication
overhead, but also is computationally very intensive. We develop efficient customized algorithms
for {\bf C1)} by exploring the structure of the SOCP and utilizing the
Alternating Direction Method of Multipliers (ADMM)
\cite{Bertsekas97,BoydADMMsurvey2011}. The main strength of our
approach is that each of its step is simple, closed-form and can be
distributed to the BSs. For the special case of the classical power
minimization problem of \cite{Wiesel06,Dahrouj10}, the new proposed
algorithm is computationally more efficient than the
existing approaches including those based on uplink-downlink duality
\cite{Rashid-Farrokhi98,Wiesel06,Dahrouj10}, and those based on the ADMM
algorithm \cite{Joshi12,Shen12} by computational complexity analysis.
Our second contribution is concerned with problem {\bf C2)}.
Specifically, we propose a novel single-stage formulation
which trades spectrum efficiency with the number of active
BSs. An efficient algorithm based on the weighted minimum mean
square error (WMMSE) algorithm \cite{Shi11} is then devised to
compute a stationary solution for the proposed problem. Once again,
this algorithm can be solved distributively among different BSs.

{\it Notations}: Boldfaced lowercase (resp.\ uppercase) letters are
used to represent vectors (resp.\ matrices). The notation ${\bf I}$ denotes the
identity matrix, and ${\bf 0}$ denotes a zero vector or matrix. The
superscripts `$H$' stands for the conjugate transpose. The set of
all $n$-dimensional real and complex vectors are denoted by $\Ral^n$
and $\Cplx^n$ respectively. The set of all real and complex
$m$-by-$n$ matrices are denoted by $\Ral^{m\times n}$ and
$\Cplx^{m\times n}$, respectively.

\section{Signal Model and Problem Statement}
\label{sec:sysmodel} %{\color{red} [[[I will suggest you think again
%whether a section for signal model is needed. It seems to me that
%writing separate subsections for both problems may be more
%reasonable (in the second section, you do not need to re-state the
%entire model, but only minor changes are needed). After all these
%two problems are not considering the same physical layer setting,
%and the last part of this section stating that because of the
%problem is NP hard, we will then consider multiple receive antennas
%is very odd to me. ]]]} {\color{blue}[[I think we can move the
%statement about NP hard in the end of this section to sec IV, and
%follows the original lines saying we would like to change "h" to "H"
%to MIMO scenario.]]}
Consider a MISO downlink multi-cell HetNet consisting of a set
$\mathcal{K}\triangleq \{1,\ldots,K\}$ of cells. Within each cell
$k$, there is a set $\mathcal{Q}_{k}=\{1,\ldots,Q_{k}\}$ distributed
base stations (BSs) which provide service to users located in
different areas of the cell. Denote
$\mathcal{Q}=\bigcup_{i=1}^{K}\mathcal{Q}_{k}$ as the set of all
BSs. Assume that in each cell $k$, a central controller has the
knowledge of all the users' data as well as their channel state
information (CSI). Its objective is to determine the transmit
beamforming vectors for all BSs within the cell. Let
$\mathcal{I}_{k}\triangleq\{1,\ldots,I_{k}\}$ denote the users
located in cell $k$, and let
$\mathcal{I}\triangleq\bigcup_{k=1}^{K}\mathcal{I}_{k}$ denote the
set of all users. Each user $i_{k}\in\mathcal{I}$ is served jointly
by a subset of BSs in $\mathcal{Q}_{k}$. For simplicity of
notations, let us assume that each BS has $M$ transmit antennas.

Let us denote ${\bf v}_{i_{k}}^{q_{k}}\in\Cplx^{M}$ as the transmit
beaformer of BS $q_{k}$ to user $i_{k}$. Define ${\bf v}\triangleq \{{\bf
v}_{i_{k}}^{q_{k}}|i_{k}\in\mathcal{I}_{k},q_{k}\in\mathcal{Q}_{k},k\in\mathcal{K}\}$
and ${\bf v}^{q_{k}}\triangleq$ $[({\bf
v}_{1_{k}}^{q_{k}})^{H},({\bf v}_{2_{k}}^{q_{k}})^{H},\ldots,({\bf
v}_{I_{k}}^{q_{k}})^{H}]^{H}$ respectively as the collection of all
the beamformers (BF) in the network, and the BFs used by BS $q_{k}$. The virtual
BF for user $i_{k}$, which consists of all the BFs that serve user
$i_k$, is given by ${\bf v}_{i_{k}}\triangleq[({\bf
v}_{i_{k}}^{1_{k}})^{H},({\bf v}_{i_{k}}^{2_{k}})^{H},\ldots,({\bf
v}_{i_{k}}^{Q_{k}})^{H}]^{H}$. Let $s_{i_{k}}\in\Cplx$ denote the
unit variance transmitted data for user $i_{k}$, then the
transmitted signal of BS $q_{k}$ can be expressed as
\begin{align} {\xb}^{q_{k}}=\sum_{i_{k}\in\mathcal{I}_{k}}{\bf
v}_{i_{k}}^{q_{k}}s_{i_{k}}.
\end{align}
The corresponding received signal of user $i_{k}$ is given by
\begin{align}
{\bf y}_{i_{k}}=\sum_{l\in\mathcal{K}}({\hb}_{i_{k}}^{l})^{H}{\bf
x}^{l}+{\bf z}_{i_{k}},
\end{align}
where $\hb_{i_{k}}^{q_{l}}\in\Cplx^{M}$ denotes the channel vector
between the BS $q_{l}$ to user $i_{k}$;
$\hb_{i_{k}}^{l}\triangleq\left[(\hb_{i_{k}}^{1_{l}})^{H},\ldots,(\hb_{i_{k}}^{Q_{l}})^{H}\right]^{H}\in
\Cplx^{MQ_{l}}$ denotes the channel matrix between $l$th cell to
user $i_k$; ${\bf x}^{k}\in\Cplx^{MQ_{k}}$ is the stacked
transmitted signal $[({\bf x}^{1_{k}})^{H},\ldots,({\bf
x}^{Q_{k}})^{H}]^{H}$ of all BSs in the $k$th cell; ${\bf
z}_{i_{k}}\in\Cplx\sim CN(0,\sigma_{i_{k}}^{2})$ is the additive
white Gaussian noise (AWGN) at user $i_{k}$. Assuming that each user
treats the interference as noise, then the
signal-to-interference-and-noise ratio (SINR) measured at the user
$i_{k}$ can be expressed as
\begin{align}
{\rm
SINR}_{i_{k}}=\frac{|\vb_{i_{k}}^{H}\hb_{i_{k}}^{k}|^{2}}{\sigma_{i_{k}}^{2}+\sum_{(l,j)\neq(k,i)}|\vb_{j_{l}}^{H}\hb_{i_{k}}^{l}|^{2}}
\end{align}
The achievable rate of user $i_{k}$ can be expressed as
\begin{align*}
R_{i_{k}}({\bf v})=\log\bigg(1+{\rm SINR}_{i_{k}}\bigg).
\end{align*}

In this work, our objective is to activate a small number of BSs to
support efficient utilization of the system resource. Such resource
utilization is measured by either one of the following two criteria:
{\bf C1)} total transmit power consumption; {\bf C2)} the overall spectrum
efficiency. Ignoring the BS activation problem for now, the BF
design problem that achieves the minimum power consumption subject
to QoS constraint can be formulated as the following SOCP
\cite{Wiesel06}
\begin{align}\label{PowerMin}
\min_{\{\vb^{q_{k}}\}}~&\sum_{q_{k}\in\mathcal{Q}}\|\vb^{q_{k}}\|_{2}^{2}\nonumber\\
\rm{s. t.}~&\|\vb^{q_{k}}\|_2^2\le P_{q_{k}},~~\forall ~q_{k}\in\mathcal{Q}\nonumber\\
&|\vb_{i_{k}}^{H}\hb_{i_{k}}^{k}|\geq\sqrt{\tau_{i_{k}}\left(\sigma_{i_{k}}^{2}+\sum_{(l,j)
\neq(k,i)}|\vb_{j_{l}}^{H}\hb_{i_{k}}^{l}|^2\right)},\\
&{\rm Im}(\vb_{i_{k}}^{H}\hb_{i_{k}}^{k})=0,
~\forall ~i_{k}\in\mathcal{I},\nonumber
\end{align}
where $\tau_{i_{k}}$ is the prescribed minimum SINR level for user
$i_k$; $P_{q_{k}}$ is the power budget of BS $q_{k},~~\forall
q_{k}\in\mathcal{Q}$, and ${\rm Im}$ denotes the imaginary part of a complex number.
It turns out that this problem is convex thus
can be solved to global optimality \cite{Wiesel06} in polynomial time.

A related BF design problem that achieves the maximum
spectrum efficiency can be formulated as the following sum rate
maximization problem
\begin{align}\label{SumRateMax}
\max_{\bf v}~&
\sum_{k\in\mathcal{K}}\sum_{i_{k}\in\mathcal{I}_{k}}R_{i_{k}}({\bf
v})\\
{\rm s. t.}~&({\bf v}^{q_{k}})^{H}{\bf v}^{q_{k}}\leq
P_{q_{k}},~~\forall
q_{k}\in\mathcal{Q}.\nonumber
\end{align}
Unfortunately, it is well-known that problem \eqref{SumRateMax} is
strongly NP-hard in general, thus it is not possible to compute its global
optimal solution in polynomial time \cite{Luo08}.

%In Sec. \ref{Sec:SumRate}, we would not restrict ourselves to
%consider single antenna users for sum rate maximization design
%criterion, but extending the number of antennas at each user to be
%multiple.

In the following sections, we will generalize problems \eqref{PowerMin}
and \eqref{SumRateMax} by incorporating nonsmooth
sparsity regularizers for BS activation, and then develop algorithms that can effectively solve the
new formulations.

\section{Base Station Activation for Power Minimization} \label{Sec:PowerMin}

%In this section, we first formulate the BS activation problem as one
%combinatorial optimization problem, and then show the formulated
%problem is NP-hard in general. Thus, we will propose an efficient
%algorithm to solve the relaxed version of the proposed formulation
%that jointly designs the beamformer while activating the smallest
%number of BSs as well as uses the minimum transmit power to support
%all the users.
%\subsection{System Model}
%{\color{blue}[[I don't think we should add one more subsection
%stating the system model. The original organization is preferred by
%me.]]}

\subsection{The Complexity for BS Activation}

Suppose all the BSs are activated, then finding the minimum transmit
power that satisfies a given QoS requirement can be formulated in
\eqref{PowerMin}. We are interested in further requiring that the
QoS targets are supported by {\it the minimum number of BSs}. A
natural two-stage approach is to first find the smallest set of BSs
that can support the QoS requirements, followed by solving problem
\eqref{PowerMin} using the set of selected BSs. In particular, the
first stage problem is given by
\begin{align}\label{PowerMinBSSelection}
\min_{\{\vb^{q_{k}}\}}~&\|\{\|\vb^{q_{k}}\|_2\}_{q_{k}\in\mathcal{Q}}\|_0  \notag\\
\rm{s. t.}~&\|\vb^{q_{k}}\|_2^2\le P_{q_{k}},~~\forall ~q_{k}\in\mathcal{Q}\notag\\
&{|\vb_{i_{k}}^{H}\hb_{i_{k}}^{k}|}\geq\sqrt{\tau_{i_{k}}\left(\sigma_{i_{k}}^{2}+\sum_{(l,j)
\neq(k,i)}|\vb_{j_{l}}^{H}\hb_{i_{k}}^{l}|^2\right)},\\
&{\rm Im}(\vb_{i_{k}}^{H}\hb_{i_{k}}^{k})=0,\quad
~\forall ~i_{k}\in\mathcal{I}\notag
\end{align}
where the $\ell_0$-norm $\|\xb\|_{0}$ denotes the number of nonzeros
elements in a vector $\xb$.

%the a SOCP problem expressed in \eqref{PowerMin} \cite{Wiesel06}.
%\begin{align}\label{PowerMin2}
%\min_{\{\vb^{q_{k}}\}}~&\sum_{q_{k}\in\mathcal{Q}}\|\vb^{q_{k}}\|_{2}^{2}\nonumber\\
%\rm{s. t.}~&\|\vb^{q_{k}}\|_2^2\le P_{q_{k}},~~\forall ~q_{k}\in\mathcal{Q}\nonumber\\
%&|\vb_{i_{k}}^{H}\hb_{i_{k}}^{k}|\geq\sqrt{\tau_{i_{k}}\left(\sigma_{i_{k}}^{2}+\sum_{(l,j)
%\neq(k,i)}|\vb_{j_{l}}^{H}\hb_{i_{k}}^{l}|^2\right)},\\
%&{\rm Im}(\vb_{i_{k}}^{H}\hb_{i_{k}}^{k})=0,
%~\forall ~i_{k}\in\mathcal{I},\nonumber
%\end{align}
%where $\tau_{i_{k}}$ is the prescribed minimum SINR level for user
%$i_k$; $P_{q_{k}}$ is the power budget of BS $q_{k},~~\forall
%q_{k}\in\mathcal{Q}$.
%Problem \eqref{PowerMin} admits a convex representation, thus its
%global optimal solution can be computed in polynomial time.

It turns out that this two-stage approach can be reformulated into a
{\it single-stage problem} shown below
\begin{align}\label{PowerMinSingleStage}
\min_{\{\vb^{q_{k}}\}}~&\|\{\|\vb^{q_{k}}\|_2\}_{q_{k}\in\mathcal{Q}}\|_0+\theta\sum_{q_{k}\in\mathcal{Q}}\|\vb^{q_{k}}\|_{2}^{2}\notag\\
\rm{s. t.}~&\|\vb^{q_{k}}\|_2^2\le P_{q_{k}},~~\forall ~q_{k}\in\mathcal{Q}\notag\\
&{|\vb_{i_{k}}^{H}\hb_{i_{k}}^{k}|}\geq\sqrt{\tau_{i_{k}}\left(\sigma_{i_{k}}^{2}+\sum_{(l,j)
\neq(k,i)}|\vb_{j_{l}}^{H}\hb_{i_{k}}^{l}|^2\right)},\\
&{\rm Im}(\vb_{i_{k}}^{H}\hb_{i_{k}}^{k})=0,
~\forall ~i_{k}\in\mathcal{I},\notag
\end{align}
where $\theta:=\frac{1}{\sum_{q_{k}\in\mathcal{Q}}P_{q_{k}}}$. The
following lemma establishes the relationship among problem
\eqref{PowerMinSingleStage}, \eqref{PowerMinBSSelection} and
\eqref{PowerMin}.
\begin{lemma}
{\it The optimal objective value of problem
\eqref{PowerMinSingleStage} lies in $[S,~S+1)$ if and only if the
optimal objective value of problem \eqref{PowerMinBSSelection} is
$S$. Furthermore, among all solutions with the optimal active BS
size equal to $S$, solving problem \eqref{PowerMinSingleStage} gives
the minimum power solution.}
\end{lemma}
$\it Proof$ Suppose $\vb^{\star}$ is an optimal solution of problem
\eqref{PowerMinBSSelection}, which yields the optimal objective
value $S$. Then the objective value of problem
\eqref{PowerMinSingleStage} is
$S+\frac{1}{\sum_{q_{k}\in\mathcal{Q}}P_{q_{k}}}\sum_{q_{k}\in\mathcal{Q}}\|\vb^{q_{k},\star}\|_2^2\in[S,~S+1)$.
On the other hand, suppose $\vb$ is optimal for problem
\eqref{PowerMinSingleStage} that achieves an objective within the
interval $[S,~S+1)$. Then the optimal solution for
\eqref{PowerMinBSSelection} cannot be smaller than $S$. Suppose the
contrary, that $\vb^{\star}$ satisfies
$\|\{\|\vb^{q_{k}\star}\|_2\}_{q_{k}\in\mathcal{Q}}\|_0\le S-1$.
Then we have
$$\|\{\|\vb^{q_{k}\star}\|_2\}_{q_{k}\in\mathcal{Q}}\|_0+\theta\sum_{q_{k}}\|\vb^{q_{k}\star}\|_2^2\le
-1+S+\theta\sum_{q_{k}}\|\vb^{q_{k}\star}\|_2^2<S,$$ which
contradicts the optimality of $\vb$. The last claim is also easy to
see by a contradiction argument. \hfill \ensuremath{\Box}

Unfortunately, despite the fact that solving the power minimization
problem \eqref{PowerMin} is easy, finding the minimum power {\it
and} the minimum number of BSs for a given set of QoS targets turns
out to be difficult. The following result makes this claim precise.
We refer the readers to Appendix \ref{subsec:NPHardProof} for the
proof.
\begin{Theorem}\label{NPHard}
{\it Solving problem \eqref{PowerMinSingleStage} is strongly NP-hard in the
number of BSs, for all $M\ge 1$. }
\end{Theorem}
%We should also mention that the NP-hardness of problem
%\eqref{PowerMinBSSelection} does not due to the existence of
%multiple transmit antennas in the BSs or there are more than one
%cells. The NP-hardness still holds true even when there is only
%single antenna at each BS and only one cell scenario is considered.
%Therefore, although we can achieve our goal by applying a two-stage
%procedure, i.e., BS selection stage by solving problem
%\eqref{PowerMinBSSelection} and then the power minimization stage
%(problem \eqref{PowerMin}) with the activated BSs from the first
%stage, there is no efficient algorithm for the first stage.

%Therefore, rather than using a two-stage procedure, we will consider
%the single-stage approach by solving problem
%\eqref{PowerMinSingleStage}.
%
%A closer look at problem \eqref{PowerMinSingleStage} reveals that
%its difficulty can be mainly attributed to the use of nonconvex
%$\ell_0$ cost function.

Motivated by the above NP-hardness result, we proceed to design
low-complexity algorithms that can obtain high-quality solutions for
problem \eqref{PowerMinSingleStage}. To this end, we propose to use
a popular relaxation scheme for this type of $\ell_0$-norm
minimization problems (e.g., \cite{Yuan06}), which replaces the
nonconvex $\ell_{0}$-norm by the $\ell_{1}$-norm. The relaxed
version of the single-stage problem \eqref{PowerMinSingleStage} can
be expressed as
\begin{center}
\fbox{\parbox[]{0.8\linewidth}{ \noindent
\begin{subequations}\label{PowerMinSingleStageRelax}
\begin{align}
f^{\min}(\vb)=\min_{\{\vb^{q_{k}}\}}~&\sum_{q_{k}\in\mathcal{Q}}\beta_{q_{k}}\|\vb^{q_{k}}\|_{2}+\theta\sum_{q_{k}\in\mathcal{Q}}\|\vb^{q_{k}}\|_{2}^{2}\\
\rm{s. t.}~&\|\vb^{q_{k}}\|_2^2\le P_{q_{k}},~~\forall ~q_{k}\in\mathcal{Q}\\
&|\vb_{i_{k}}^{H}\hb_{i_{k}}^{k}|\geq\sqrt{\tau_{i_{k}}\left(\sigma_{i_{k}}^{2}+\sum_{(l,j)
\neq(k,i)}|\vb_{j_{l}}^{H}\hb_{i_{k}}^{l}|^2\right)},\\
&{\rm Im}(\vb_{i_{k}}^{H}\hb_{i_{k}}^{k})=0,
~\forall ~i_{k}\in\mathcal{I},
\end{align}
\end{subequations}
 }}
\end{center}
where $\beta_{q_{k}}\in\Ral,~~\forall  q_{k}\in\mathcal{Q}$ are given
parameters to control the number of active BSs of the obtained
solution of problem \eqref{PowerMinSingleStageRelax}. In Sec.
\ref{sec:reweight}, we will further discuss how these parameters
can be adaptively chosen. Since problem
\eqref{PowerMinSingleStageRelax} is a SOCP (just like
problem \eqref{PowerMin}), it can be solved to global optimality using a standard
package such as CVX \cite{cvx}. {However, using general purpose solvers
can be slow, especially when the number of variables
$\sum_{k\in\mathcal{K}}M Q_{k} I_{k}$ and the number of constraints
$2|\mathcal{I}|+|\mathcal{Q}|$ become large.}
%Furthermore, directly solving problem
%\eqref{PowerMinSingleStageRelax} requires a controller to solve it
%centrally.
In what follows, we will exploit the structure of the problem at
hand, and develop a fast distributed algorithm for solving
problem \eqref{PowerMinSingleStageRelax}. Our approach is based on
the well-known ADMM algorithm \cite{BoydADMMsurvey2011}, which we
outline briefly below.

\subsection{A Brief Review of the ADMM Algorithm}

The ADMM algorithm was originally developed in 1970s, and has
attracted lots of interests recently due to its efficiency in
large-scale optimization (see \cite{BoydADMMsurvey2011} and references
therein). Specifically, the ADMM is designed to solve the following
structured convex problem
\begin{align}\label{ADMM}
\min_{{\bf x}\in\Cplx^{n},{\bf z}\in\Cplx^{m}}~& f({\bf x})+g({\bf z})\notag\\
{\rm s.t.}~& {\bf A}{\bf x}+{\bf B}{\bf z}={\bf c}\\
& {\bf x}\in \mathcal{C}_{1},~{\bf z}\in\mathcal{C}_{2}\notag
\end{align}
where ${\bf A}\in\Cplx^{k\times n}$, ${\bf B}\in\Cplx^{k\times m}$,
${\bf c}\in\Cplx^{k}$, and $f$ and $g$ are convex functions while
$\mathcal{C}_{1}$ and $\mathcal{C}_{2}$ are non-empty convex
sets. %An equivalent formulation for problem \eqref{ADMM} is to add
%to its objective the augmented Lagrangian penalization
%$(\rho/2)\|{\bf A}{\bf x}+{\bf B}{\bf z}-{\bf z}\|_2^2$, that is
%\begin{align}\label{PenalizedLag}
%\min~& f({\bf x})+g({\bf z})+(\rho/2)\|{\bf A}{\bf x}+{\bf B}{\bf z}-{\bf c}\|_2^2\notag\\
%{\rm s.t.}~& {\bf A}{\bf x}+{\bf B}{\bf z}={\bf c}\\
%& {\bf x}\in \mathcal{C}_{1},~{\bf z}\in\mathcal{C}_{2}.\notag
%\end{align}
%where $\rho>0$ is some constant.
The partial augmented Lagrangian function for problem \eqref{ADMM}
can be expressed as
\begin{align}
L_{\rho}({\bf x},{\bf z}, {\bf y})=f({\bf x})+g({\bf z})+{\rm
Re}\left({\bf y}^{H}({\bf A}{\bf x}+{\bf B}{\bf z}-{\bf
c})\right)+(\rho/2)\|{\bf A}{\bf x}+{\bf B}{\bf z}-{\bf c}\|_2^2
\end{align}
where ${\bf y}\in\Cplx^{k}$ is the Lagrangian dual variables
associated with the linear equality constraint, and $\rho>0$ is some
constant. The ADMM algorithm solves problem \eqref{ADMM} by
iteratively performing three steps in each iteration $t$:
\begin{subequations}
\begin{align}
{\bf x}^{(t)}&=\arg\min_{{\bf x}}L_{\rho}({\bf x},{\bf
z}^{(t-1)},{\bf
y}^{(t-1)})\label{ADMMUpdate1}\\
{\bf z}^{(t)}&=\arg\min_{{\bf z}}L_{\rho}({\bf x}^{(t)},{\bf z},{\bf
y}^{(t-1)})\label{ADMMUpdate2}
\\
{\bf y}^{(t)}&={\bf y}^{(t)}+\rho({\bf A}{\bf x}^{(t)}+{\bf B}{\bf
z}^{(t)}-{\bf c})\label{ADMMUpdate3}.
\end{align}
\end{subequations}
The efficiency of ADMM mainly comes from the fact in many
applications, the subproblems for the primal variables
\eqref{ADMMUpdate1} and \eqref{ADMMUpdate2} can be solved easily in
closed-form. The convergence property of this algorithm
is summarized in the following lemma.
\begin{proposition}\label{ADMMConverge}\cite{Bertsekas97} Assume that the optimal solution set of problem \eqref{ADMM} is non-empty, and ${\bf A}^{T}{\bf A}$ and ${\bf B}^{T}{\bf B}$ are invertible. Then the sequence of $\{{\bf x}^{k}, {\bf z}^{k},{\bf y}^{k}\}$ generated by \eqref{ADMMUpdate1}, \eqref{ADMMUpdate2}, and \eqref{ADMMUpdate3} is bounded and every limit point of $\{{\bf x}^{k},{\bf z}^{k}\}$ is an
optimal solution of problem \eqref{ADMM}.
\end{proposition}

\subsection{The Proposed ADMM Approach}

In this subsection, we will show that our joint BS activation and
power minimization problem \eqref{PowerMinSingleStageRelax} can be
in fact solved very efficiently by using the ADMM.

%Due to the fact that QoS constraints of problem
%\eqref{PowerMinSingleStageRelax} require each cell jointly optimize
%the beamformers among each cell, this problem should be solved
%centrally. However, if for every pair of users $i_{k}$ and
%$j_{l}\in\mathcal{I}$, the interference to user $j_{l}$ from signal
%for user $i_{k}$, expressed as
%$K_{j_{l}}^{i_{k}}=(\hb_{j_{l}}^{k})^{H}\vb_{i_{k}}$ are given and
%fixed at both cells $k$ and $l\in\mathcal{K}$, the QoS constraints
%of problem \eqref{PowerMinSingleStageRelax} can be separable over
%cells. On the other hand, to exploit the structure of the objective
%function of problem \eqref{PowerMinSingleStageRelax}, we will
%introduce the copy of each beamformer $\vb^{q_{k}}$ by
%$\wb^{q_{k}}=\vb^{q_{k}},~~\forall  q_{k}\in\mathcal{Q}$, and the
%power budget constraint is only on the $\wb^{q_{k}}$. Thus, the
%objective function can be separable among $\wb^{q_{k}}$ and
%$\vb^{q_{k}}$. Since each $\wb^{q_{k}}$ relates to $\vb^{q_{k}}$ by
%a linear constraint as each $K_{j_{l}}^{i_{k}},~~\forall
%i_{k},j_{l}\in\mathcal{I}$, does.

The main idea is to decompose the tightly coupled network problem
into several subproblems of much smaller sizes, each of which can be
solved in closed form. For example, by introducing a copy
$\wb^{q_k}$ for the original BF $\vb^{q_k}$, the objective function
of problem \eqref{PowerMinSingleStageRelax} can be separated into
two parts
\begin{align}
&\sum_{q_{k}\in\mathcal{Q}}\beta_{q_{k}}\|\wb^{q_{k}}\|_{2}+\theta\sum_{q_{k}\in\mathcal{Q}}\|\vb^{q_{k}}\|_{2}^{2},
\end{align}
where each part is further separable among the BSs. In this way,
after some further manipulation which will be shown shortly, it
turns out that solving the subproblem for either $\wb$ or $\vb$ can
be made very easy.

Formally, let us introduce a few new variables
\begin{subequations}\label{NewVariables}
\begin{align}
K_{j_{l}}^{i_{k}}:&=(\hb_{j_{l}}^{k})^{H}\vb_{i_{k}},~\forall ~i_{k},j_{l}\in\mathcal{I},\label{NewVariablesK}\\
\wb^{q_{k}}:&=\vb^{q_{k}}, ~\forall ~q_{k}\in\mathcal{Q},\label{NewVariablesWV}\\
\kappa_{i_{k}}:&=\hat\kappa_{i_{k}}=\sigma_{i_{k}}\in\Ral,~\forall ~i_{k}\in\mathcal{I}.\label{NewVariablesKappa}
\end{align}
\end{subequations}
and define
$\Kb\triangleq\{K_{j_{l}}^{i_{k}}\mid i_{k},j_{l}\in\mathcal{I}\}$,
$\wb\triangleq\{\wb^{q_{k}}\mid q_{k}\in\mathcal{Q}\}$,
$\vb\triangleq\{\vb^{q_{k}}\mid q_{k}\in\mathcal{Q}\}$,
$\mbox{\boldmath$\kappa$}\triangleq\{\kappa_{i_{k}}\mid i_{k}\in\mathcal{I}\}$
and $\mbox{\boldmath$\hat\kappa$}\triangleq\{\hat
\kappa_{i_{k}}\mid i_{k}\in\mathcal{I}\}$. Clearly $K_{j_{l}}^{i_{k}}$
represents the interference level experienced at user $j_l$
contributed by the BF for user $i_k$; $\wb^{q_k}$ is a copy of the
original BF $\vb^{q_k}$; $\kappa_{i_k}$ and $\hat{\kappa}_{i_k}$ are
copies of the noise power $\sigma_{i_k}$. %The idea is to combine the
%set of interference $\Kb$ and the beamformer copies $\wb$ as one
%optimization variable while the original beamformer $\vb$ as another
%optimization variable.

%The new variable set
%$\mbox{\boldmath$\kappa$}\triangleq\{\kappa_{i_{k}}|i_{k}\in\mathcal{I}\}$
%and $\mbox{\boldmath$\hat\kappa$}\triangleq\{\hat
%\kappa_{i_{k}}|i_{k}\in\mathcal{I}\}$ are also introduced to make
%sure the updating steps are in closed-form as explained later. With
%these new variables,

With these new variables, problem \eqref{PowerMinSingleStageRelax}
can be equivalently expressed as
\begin{subequations}\label{PowerMinSingleStageADMM}
\begin{align}
\min_{\{\vb^{q_{k}}\},\{\wb^{q_{k}}\},\{K_{j_{l}}^{i_{k}}\},\{\kappa_{i_{k}}\},\{\hat\kappa_{i_{k}}\}}~&\sum_{q_{k}\in\mathcal{Q}}\beta_{q_{k}}\|\wb^{q_{k}}\|_{2}+\theta\sum_{q_{k}\in\mathcal{Q}}\|\vb^{q_{k}}\|_{2}^{2}\\
\rm{s. t.}~&\|\wb^{q_{k}}\|_2^2\le P_{q_{k}},~~\forall ~q_{k}\in\mathcal{Q}\\
&|K_{i_{k}}^{i_{k}}|\geq\sqrt{\tau_{i_{k}}\left(\kappa_{i_{k}}^2+\sum_{(l,j)
\neq(k,i)}|K_{i_{k}}^{j_{l}}|^2\right)},\\
&{\rm Im}(K_{i_{k}}^{i_{k}})=0, ~\forall ~i_{k}\in\mathcal{I},\\
&
\eqref{NewVariablesK},~\eqref{NewVariablesWV},~\mbox{and}~\eqref{NewVariablesKappa}
\end{align}
\end{subequations}
The partial augmented Lagrangian function of the above problem is
given by
\begin{align}
L(\wb,\Kb,\mbox{\boldmath$\kappa$},\vb,\mbox{\boldmath$\hat\kappa$},\bfmu,\bflambda,\mbox{\boldmath$\delta$})
&=\sum_{q_{k}\in\mathcal{Q}}\beta_{q_{k}}\|\wb^{q_{k}}\|_2+\theta\sum_{q_{k}\in\mathcal{Q}}\|\vb^{q_{k}}\|_2^2+\sum_{i_{k}\in\mathcal{I}}(\kappa_{i_{k}}-\hat\kappa_{i_{k}})\delta_{i_{k}}\nonumber\\
&\quad+{\rm Re}\left(\sum_{i,k,j,l}\langle
K_{j_{l}}^{i_{k}}-(\hb_{j_{l}}^{k})^{H}\vb_{i_{k}},\mu_{j_{l}}^{i_{k}}\rangle\right)+{\rm
Re}\left(\sum_{q_{k}\in\mathcal{Q}}\langle\wb^{q_{k}}-\vb^{q_{k}},\bflambda^{q_{k}}\rangle\right)+
\nonumber\\
&\quad+\frac{\rho}{2}\sum_{i_{k},j_{l}\in\mathcal{I}}\left|K_{j_{l}}^{i_{k}}-(\hb_{j_{l}}^{k})^{H}\vb_{i_{k}}\right|^2
+\frac{\rho}{2}\sum_{q_{k}\in\mathcal{Q}}\|\wb^{q_{k}}-\vb^{q_{k}}\|_2^2+\frac{\rho}{2}\sum_{i_{k}\in\mathcal{I}}
(\kappa_{i_{k}}- \hat\kappa_{i_{k}})^2,\label{eqLagrangian2}
\end{align}
where
$\bfmu\triangleq\{\mu_{j_{l}}^{i_{k}}\in{\Cplx}\mid i_{k},j_{l}\in\mathcal{I}\}$,
$\bflambda\triangleq\{\bflambda^{q_{k}}\in{\Cplx}^{I_{k}}\mid q_{k}\in\mathcal{Q}\}$,
and
$\mbox{\boldmath$\delta$}\triangleq\{\delta_{i_{k}}\in\Ral\mid i_{k}\in\mathcal{I}\}$
are, respectively, the Lagrangian dual variable for constraints
\eqref{NewVariablesK}, \eqref{NewVariablesWV}, and
\eqref{NewVariablesKappa}.

It can be readily observed that problem
\eqref{PowerMinSingleStageADMM} is separable among the block
variables $\{\vb,\mbox{\boldmath$\hat\kappa$}\}$ and
$\{\wb,\Kb,\mbox{\boldmath$\kappa$}\}$. Moreover, all the
constraints linking these two block of variables (i.e.,
\eqref{NewVariablesK}, \eqref{NewVariablesWV}, and
\eqref{NewVariablesKappa}) are linear equalities. Therefore, ADMM
algorithm can be directly applied to solve problem
\eqref{PowerMinSingleStageADMM}. The main algorithmic steps are
summarized in Algorithm 1.

\begin{center}
\fbox{\parbox[]{.99\linewidth}{ \noindent Algorithm 1: ADMM for
\eqref{PowerMinSingleStageRelax}:
\begin{algorithmic}[1]
\State {\bf Initialize} all primal variables
$\wb^{(0)},\vb^{(0)},\Kb^{(0)}$ (do not need to be feasible for
problem \eqref{PowerMinSingleStageADMM}); Initialize all dual
variables $\bfmu^{(0)},\bflambda^{(0)}$;

\State {\bf Repeat}

\State ~~~Solve the following problem and obtain $\{\wb^{(t+1)},\Kb^{(t+1)},\mbox{\boldmath$\kappa$}^{(t+1)}\}$ (\eqref{ProblemKSOCP}, \eqref{OptimalSolutionW})
\begin{align*}
\min_{\wb,\Kb,\mbox{\boldmath$\kappa$}}~&
L(\wb,\Kb,\mbox{\boldmath$\kappa$},\vb^{(t)},\mbox{\boldmath$\hat\kappa$}^{(t)},\bfmu^{(t)},\bflambda^{(t)},\mbox{\boldmath$\delta$}^{(t)})\\
\rm{s. t.}~&\|\wb^{q_{k}}\|_2^2\le P_{q_{k}},~~\forall
q_{k}\in\mathcal{Q}\\
&|K_{i_{k}}^{i_{k}}|\geq\sqrt{\tau_{i_{k}}\left(\kappa_{i_{k}}^{2}+\sum_{(l,j)
\neq(k,i)}|K_{i_{k}}^{j_{l}}|^2\right)},~ {\rm Im}(K_{i_{k}}^{i_{k}})=0, ~\forall ~i_{k}\in\mathcal{I};
\end{align*}

\State ~~~Solve the following problem and obtain
$\vb^{(t+1)},\mbox{\boldmath$\hat\kappa$}^{(t+1)}$
(\eqref{OptimalSolutionV})
\begin{align*}
\min_{\vb, \mbox{\boldmath$\hat\kappa$}}~&
L(\wb^{(t+1)},\Kb^{(t+1)},\mbox{\boldmath$\kappa$}^{(t+1)},\vb,\mbox{\boldmath$\hat\kappa$},\bfmu^{(t)},\bflambda^{(t)},\mbox{\boldmath$\delta$}^{(t)})\\
\rm{s. t.}~&\hat\kappa_{i_{k}}=\sigma_{i_{k}},~~\forall
i_{k}\in\mathcal{I};
\end{align*}

\State ~~~Update the multipliers by
\begin{align*}
\mu_{j_{l}}^{i_{k}(t+1)}&=\mu_{j_{l}}^{i_{k}(t)}+\rho\left(K_{j_{l}}^{i_{k}(t+1)}-(\hb_{j_{l}}^{k})^{H}\vb_{i_{k}}^{(t+1)}\right),~~\forall ~i_{k},j_{l}\in\mathcal{I}\\
\bflambda^{q_{k}(t+1)}&=\bflambda^{q_{k}(t)}+\rho(\wb^{q_{k}(t+1)}-\vb^{q_{k}(t+1)}),~~\forall
q_{k}\in\mathcal{Q}_{k}\\
\delta_{i_{k}}^{(t+1)}&=\delta_{i_{k}}^{(t)}+\rho(\kappa_{i_{k}}^{(t+1)}-\hat\kappa_{i_{k}}^{(t+1)}),~~\forall
q_{k}\in\mathcal{Q}_{k};
\end{align*}

\State{\bf Until} Desired stopping criteria is met

\end{algorithmic}
}}
\end{center}

Before further investigating how each update procedure can be solved
in closed-form, let us first discuss the convergence result for the
proposed algorithm.
\begin{Theorem}\label{ConvergenceADMM}
Assume that problem \eqref{PowerMinSingleStageRelax} is feasible.
Every limit point $\vb^{(t)}$ (or $\wb^{(t)}$) generated by
Algorithm~1 is an optimal solution of problem
\eqref{PowerMinSingleStageRelax}.
\end{Theorem}
$\it Proof.$ {Let us stack all elements of
$\{\wb,\Kb,\mbox{\boldmath$\kappa$}\}$ and
$\{\vb,\mbox{\boldmath$\hat\kappa$}\}$ to vectors
$\{\wb_{stack}\in\Cplx^{M|\mathcal{Q}||\mathcal{I}|},\Kb_{stack}\in\Cplx^{|\mathcal{I}|^{2}},\mbox{\boldmath$\kappa$}_{stack}\in\Ral^{|\mathcal{I}|}\}$
and
$\{\vb_{stack}\in\Cplx^{M|\mathcal{Q}||\mathcal{I}|},\mbox{\boldmath$\hat\kappa$}_{stack}\in\Ral^{|\mathcal{I}|}\}$. Then, by
comparing problem \eqref{ADMM} and problem
\eqref{PowerMinSingleStageADMM}, when
$\xb=[\wb_{stack}^{H},\Kb_{stack}^{H},\mbox{\boldmath$\kappa$}_{stack}^{H}]^{H}$
and
$\zb=[\vb_{stack}^{H},\mbox{\boldmath$\hat\kappa$}_{stack}^{H}]^{H}$
we can observe that
\begin{align*}
f(\xb)&=
\sum_{q_{k}\in\mathcal{Q}}\beta_{q_{k}}\|\wb^{q_{k}}\|_{2},~
g(\zb)=\theta\sum_{q_{k}\in\mathcal{Q}}\|\vb^{q_{k}}\|_{2}^{2},~\Ab=\Ib,~{\bf B}=-\left[\begin{array}{cc}\Ib&{\bf 0}\\
\Hb_{stack} & {\bf
0}\\
{\bf 0} & \Ib\end{array}\right],~\cb={\bf 0}\\
\mathcal{C}_{1}&= \Bigg\{\xb\mid \|\wb^{q_{k}}\|_2^2\le
P_{q_{k}},~~\forall ~q_{k}\in\mathcal{Q},\\
&\quad\quad|K_{i_{k}}^{i_{k}}|\geq\sqrt{\tau_{i_{k}}\left(\kappa_{i_{k}}^2+\sum_{(l,j)
\neq(k,i)}|K_{i_{k}}^{j_{l}}|^2\right)}, {\rm
Im}(K_{i_{k}}^{i_{k}})=0, ~\forall ~i_{k}\in\mathcal{I},\Bigg\},\\
\mathcal{C}_{2}&=\{\zb\mid \hat \kappa_{i_{k}}=\sigma_{i_{k}},~~\forall
i_{k}\in\mathcal{I}\},
\end{align*}
where $\Hb_{stack}\in\Cplx^{|\mathcal{I}|^{2}\times
M|\mathcal{Q}||\mathcal{I}|}$ is a stacked matrix of
$\{(\hb_{j_{l}}^{k})^{H}\mid j_{l}\in\mathcal{I},k\in\mathcal{K}\}$ and
${\bf 0}$'s in a way that ${\bf
K}_{stack}-\Hb_{stack}\vb_{stack}=\mathbf{0}$ is equivalent to
\eqref{NewVariablesK}}.

Since $\Ab^{T}\Ab=\Ib$ and $\Bb^{T}\Bb=\left[\begin{array}{cc}\Ib+\Hb_{stack}^{T}\Hb_{stack} & {\bf 0}\\
{\bf 0} &\Ib\end{array}\right]$ are invertible, and both ${\cal{C}}_1$ and
${\cal{C}}_2$ are convex sets, then by Proposition~\ref{ADMMConverge}, we
can conclude that every limit point $\vb^{(t)}$ (or $\wb^{(t)}$) of
Algorithm 1 is an optimal solution of problem
\eqref{PowerMinSingleStageRelax}. \hfill \ensuremath{\Box}

\subsection{Step-by-Step Computation for the Proposed Algorithm}
In the following, we will explain in detail how each primal
variables $\wb, \Kb, \mbox{\boldmath$\kappa$}, \vb$, and
$\mbox{\boldmath$\hat\kappa$}$ (ignoring the superscript iteration
index for simplicity) is updated. As will be seen shortly, the
update for the first block  $\{\wb, \Kb, \mbox{\boldmath$\kappa$}\}$
can be further decomposed into two independent problems, one for
$\wb$, and one for $\{\Kb, \mbox{\boldmath$\kappa$}\}$.

%Moreover, for the updating step
%for block $\{\wb, \Kb, \mbox{\boldmath$\kappa$}\}$, i.e., line 3 of
%the proposed ADMM approach, due to the constraints and the augmented
%Lagrange function are separable between $\{\Kb,
%\mbox{\boldmath$\kappa$}\}$ and $\wb$, this updating step can be
%conducted separately between these two variable blocks.

{\bf (1) Update $\{\Kb, \mbox{\boldmath$\kappa$}\}$}: First observe
that the subproblem related to $\{\Kb, \mbox{\boldmath$\kappa$}\}$
is independent of $\wb$, and can be decoupled over each user.
Therefore it can be written as $|\mathcal{I}|$ separate problems,
with $i_{k}$-th subproblem expressed as
\begin{align}
\min_{\{K_{i_{k}}^{j_{l}}\}_{j_{l}\in\mathcal{I}},\kappa_{i_{k}}}~&{\rm
Re}\left(\sum_{i_{k},j_{l}\in\mathcal{I}}\langle
K_{j_{l}}^{i_{k}}-(\hb_{j_{l}}^{k})^{H}\vb_{i_{k}},\mu_{j_{l}}^{i_{k}}\rangle\right)+\delta_{i_{k}}(\kappa_{i_{k}}-\hat\kappa_{i_{k}})\nonumber\\
&+\frac{\rho}{2}\sum_{j_{l}\in\mathcal{I}}\left|K_{i_{k}}^{j_{l}}-(\hb_{i_{k}}^{l})^{H}\vb_{j_{l}}\right|^2
+\frac{\rho}{2}(\kappa_{i_{k}}-\hat\kappa_{i_{k}})^2\nonumber\\
\rm{s. t.}&\quad
|K_{i_{k}}^{i_{k}}|\geq\sqrt{\tau_{i_{k}}\left(\kappa_{i_{k}}^{2}+\sum_{(l,j)
\neq(k,i)}|K_{i_{k}}^{j_{l}}|^2\right)},\label{ProblemK}\\
& \quad {\rm Im}(K_{i_{k}}^{i_{k}})=0.\nonumber
\end{align}
By completing the squares, this problem can be equivalently written
as
\begin{align}
\min_{\{K_{i_{k}}^{j_{l}}\}_{j_{l}\in\mathcal{I}},\kappa_{i_{k}}}~&\left(\kappa_{i_{k}}-\hat\kappa_{i_{k}}+\frac{\delta_{i_{k}}}{\rho}\right)^{2}
+\sum_{j_{l}\in\mathcal{I}}\left|K_{i_{k}}^{j_{l}}-(\hb_{i_{k}}^{l})^{H}\vb_{j_{l}}+\frac{\mu_{i_{k}}^{j_{l}}}{\rho}\right|^2\nonumber\\
\rm{s. t.}&\quad
|K_{i_{k}}^{i_{k}}|\geq\sqrt{\tau_{i_{k}}\left(\kappa_{i_{k}}^{2}+\sum_{(l,j)
\neq(k,i)}|K_{i_{k}}^{j_{l}}|^2\right)},\label{ProblemKSOCP}\\
& \quad {\rm Im}(K_{i_{k}}^{i_{k}})=0.\nonumber
\end{align}
Let us use
$\{\{K_{i_{k}}^{j_{l}\star}\}_{j_{l}\in\mathcal{I}},\kappa_{i_{k}}^{\star}\}$
to denote the optimal solution of problem \eqref{ProblemKSOCP}. Then
the corresponding first-order optimality conditions are given by
\begin{subequations}
\begin{align}
&  K_{i_{k}}^{i_{k}\star}=\frac{1}{2}\gamma^{\star}+{\rm Re}\left((\hb_{i_{k}}^{k})^{H}\vb_{i_{k}}-\frac{\mu_{i_{k}}^{i_{k}}}{\rho}\right)\label{OptimalK}\\
&  K_{i_{k}}^{j_{l}\star}=\frac{\bar K_{i_{k}}\left((\hb_{i_{k}}^{l})^{H}\vb_{j_{l}}-\frac{\mu_{i_{k}}^{j_{l}}}{\rho}\right)}{\bar K_{i_{k}}+\frac{\gamma^{\star}\sqrt{\tau_{i_{k}}}}{2}},~~\forall  j_{l}\in\mathcal{I},j_{l}\neq i_{k}\label{OptimalK2}\\
& \kappa_{i_{k}}^{\star}=\frac{\bar K_{i_{k}}\left(\hat\kappa_{i_{k}}-\frac{\delta_{i_{k}}}{\rho}\right)}{\bar K_{i_{k}}+\frac{\gamma^{\star}\sqrt{\tau_{i_{k}}}}{2}}\label{OptimalKappa}\\
&  K_{i_{k}}^{i_{k}\star}\geq\sqrt{\tau_{i_{k}}}\bar
K_{i_{k}},~\gamma^{\star}\geq0,~\left(K_{i_{k}}^{i_{k}\star}-\sqrt{\tau_{i_{k}}}\bar
K_{i_{k}}\right)\gamma^{\star}=0\label{ComplementarityK}
\end{align}
\end{subequations}
where $\gamma^{\star}$ is the optimal Lagrangian dual variable for
the second-order cone constraint of problem \eqref{ProblemKSOCP} and
$\bar
K_{i_{k}}\triangleq\sqrt{\kappa_{i_{k}}^{\star2}+\sum_{(l,j)\neq(k,i)}\left|K_{i_{k}}^{j_{l}\star}\right|^{2}}$.
If $\gamma^\star=0$, the objective value of problem
\eqref{ProblemKSOCP} is the minimum possible value, $0$, and by
complementarity condition \eqref{ComplementarityK}, this is possible
only if
\begin{align}\label{CheckK}
\left|{\rm
Re}\left((\hb_{i_{k}}^{k})^{H}\vb_{i_{k}}-\frac{\mu_{i_{k}}^{i_{k}}}{\rho}\right)\right|\geq
\sqrt{\tau_{i_{k}}}\sqrt{(\hat\kappa_{i_{k}}-\frac{\delta_{i_{k}}}{\rho})^2+\sum_{(l,j)\neq(k,i)}\left|(\hb_{i_{k}}^{l})^{H}\vb_{j_{l}}-\frac{\mu_{i_{k}}^{j_{l}}}{\rho}\right|^{2}}\triangleq
\underline{K}_{i_{k}}.
\end{align}
On the other hand, if \eqref{CheckK} does not hold, we know that
$\gamma^{\star}\neq 0$, and by complementarity condition
\eqref{ComplementarityK}, ${\rm
Re}(K_{i_{k}}^{i_{k}\star})=\sqrt{\tau_{i_{k}}}\bar K_{i_{k}}$
holds. Therefore, the optimal dual variable, $\gamma^{\star}$ can be
analytically solved as
\begin{align*}
\gamma^{\star}=2\frac{\underline{K}_{i_{k}}-{\rm
Re}\left((\hb_{i_{k}}^{k})^{H}\vb_{i_{k}}-\frac{\mu_{i_{k}}^{i_{k}}}{\rho}\right)}{1+\tau_{i_{k}}}
\end{align*}
Hence, the optimal solution of problem \eqref{ProblemKSOCP} can be
solved in closed-form by \eqref{OptimalK}, \eqref{OptimalK2}, and
\eqref{OptimalKappa} with given $\gamma^{\star}$ and the fact that
$\bar K_{i_{k}}={\rm
Re}(K_{i_{k}}^{i_{k}\star})/\sqrt{\tau_{i_{k}}}$.

It is worth noting that, this closed-form update rule is made
possible by making $\kappa_{i_{k}}$ as an optimization variable.
This is the reason that we want to introduce extra variables
$\{\kappa_{i_k}\}$ and $\{\hat{\kappa}_{i_k}\}$ in
\eqref{NewVariablesKappa}.

{\bf (2) Update $\{\wb\}$}: The subproblem for the optimization
variable $\wb$ can also be decoupled over $|\mathcal{Q}|$ separate
subproblems, one for each BS $q_k$:
\begin{align}
\min_{\wb^{q_{k}}}~&\beta_{q_{k}}\|\wb^{q_{k}}\|_2+\frac{\rho}{2}\|\wb^{q_{k}}-\vb^{q_{k}}-\bflambda^{q_{k}}/\rho\|_2^2\nonumber\\
\rm{s. t.}&\quad\|\wb^{q_{k}}\|_2^2\le P_{q_{k}}.\label{problemW}
\end{align}
By defining $\bb^{q_{k}}=\vb^{q_{k}}+\bflambda^{q_{k}}/\rho$, the
optimal solution $\wb^{q_{k}\star}$ should satisfy the first-order
optimality condition
\begin{subequations}
\begin{align}
&\rho\bb^{q_{k}}-\wb^{q_{k}\star}(\rho+2\gamma^{q_{k}\star})\in\partial(\beta_{q_{k}}\|\wb^{q_{k}\star}\|_2)\label{SubgradientW}\\
&\|\wb^{q_{k}\star}\|_2^{2}\leq P_{q_{k}},~\gamma^{q_{k}\star}\geq 0\\
&(\|\wb^{q_{k}\star}\|_2^{2}-P_{q_{k}})\gamma^{q_{k}\star}=0
\end{align}
\end{subequations}
where $\gamma^{q_{k}\star}$ is the optimal Lagrangian multiplier
associated with the quadratic constraint $\|\wb^{q_{k}}\|_2^2\le
P_{q_{k}}$. From \eqref{SubgradientW} and the definition of the
subgradient for the $\ell_2$ norm, we have that
$\wb^{q_{k}\star}={\bf 0}$ whenever $\rho\|\bb^{q_{k}}\|_2\leq
\beta_{q_{k}}$. When $\rho\|\bb^{q_{k}}\|_2>\beta_{q_{k}}$, we have
\begin{align}\label{eqWStar}
&\rho\bb^{q_{k}}-\wb^{q_{k}\star}(\rho+2\gamma^{q_{k}\star})=\beta_{q_{k}}\frac{\wb^{q_{k}\star}}{\|\wb^{q_{k}\star}\|_2}\nonumber\\
\Longrightarrow ~
&\wb^{q_{k}\star}=\frac{\bb^{q_{k}}(\rho\|\bb^{q_{k}}\|_2-\beta_{q_{k}})}{(\rho+2\gamma^{q_{k}\star})\|\bb^{q_{k}}\|_2}.
\end{align}
By the complementarity condition, $\gamma^{q_{k}\star}=0$ if
$\left\|\frac{\bb^{q_{k}}(\rho\|\bb^{q_{k}}\|_2-\beta_{q_{k}})}{\rho\|\bb^{q_{k}}\|_2}\right\|_2^{2}\leq
P_{q_{k}}$. Otherwise, $\gamma^{q_{k}\star}$ should be chosen such
that $\|\wb^{q_{k}\star}\|_2^2=P_{q_{k}}$, which implies that
$\gamma^{q_{k}\star}=(\rho(\|\bb^{q_{k}}\|_2-\sqrt{P_{q_{k}}})-\beta_{q_{k}})/(2\sqrt{P_{q_{k}}})$.
Plugging these choices of $\gamma^{q_{k}\star}$ into
\eqref{eqWStar}, then we conclude that the solution for problem
\eqref{problemW} is given by
\begin{align}
\wb^{q_{k}\star}=\left\{ \begin{array}{ll}
\mathbf{0},&\rho\|\bb^{q_{k}}\|\le \beta_{q_{k}},\\
\frac{\bb^{q_{k}}(\rho\|\bb^{q_{k}}\|_2-\beta_{q_{k}})}{\rho\|\bb^{q_{k}}\|_2},&
\rho\|\bb^{q_{k}}\|> \beta_{q_{k}}~\mbox{and}~
\left\|\frac{\bb^{q_{k}}(\rho\|\bb^{q_{k}}\|_2-\beta_{q_{k}})}{\rho\|\bb^{q_{k}}\|_2}\right\|_2^{2}\leq
P_{q_{k}},\\
\sqrt{P_{q_{k}}}\frac{\bb^{q_{k}}}{\|\bb^{q_{k}}\|_2},
&{\rm otherwise.}
\end{array}\right.\label{OptimalSolutionW}
\end{align}

{\bf (3) Update $\vb,\mbox{\boldmath$\hat\kappa$}$}: From step 4 of
Algorithm 1, we readily have
$\hat\kappa_{i_{k}}^{\star}={\sigma_{i_{k}}},~~\forall
i_{k}\in\mathcal{I}$. The subproblem for the block variable $\vb$
can be written as $K$ independent unconstrained quadratic problems, one for
each cell $k$:
\begin{align}
\!\!\!\!\min_{\{\vb^{q_{k}}\}_{q_{k}\in\mathcal{Q}_{k}}}&
\frac{\rho}{2}\sum_{i_{k}\in\mathcal{I}_{k}\atop j_{l}\in\mathcal{I}}\left|(\hb_{j_{l}}^{k})^{H}\vb_{i_{k}}-K_{j_{l}}^{i_{k}}-{\mu_{j_{l}}^{i_{k}}}/{\rho}\right|^{2}+\frac{\rho}{2}\sum_{q_{k}\in\mathcal{Q}_{k}}\|\vb^{q_{k}}-\wb^{q_{k}}+\bflambda^{q_{k}}/\rho\|_2^2
+\theta\sum_{q_{k}\in\mathcal{Q}_{k}}\|\vb^{q_{k}}\|_2^2.\label{problemV}
\end{align}
%The above unconstrained problem can be written equivalently as
%\begin{align}
%\min_{\{\vb_{i_{k}}\}_{i_{k}\in\mathcal{I}_{k}}}~&
%\sum_{i_{k}\in\mathcal{I}_{k}}\left(\frac{\rho}{2}\sum_{j_{l}\in\mathcal{I}}\left((\hb_{j_{l}}^{k})^{H}\vb_{i_{k}}-K_{j_{l}}^{i_{k}}-{\mu_{j_{l}}^{i_{k}}}/{\rho}\right)^{2}+\frac{\rho}{2}\|\vb_{i_{k}}-\wb_{i_{k}}+\bflambda_{i_{k}}/\rho\|_2^2
%+\theta\|\vb_{i_{k}}\|_2^2\right).
%\end{align}
The solution for this unconstrained problem is given by
\begin{align}
\vb_{i_{k}}^{\star}=\rho^{-1}\left((1+2\theta/\rho)\Ib+\Hb^{k}\Hb^{kH}\right)^{-1}(\rho\Hb^{k}
\Kb^{i_{k}}+\Hb\bfmu^{i_{k}}+\rho\wb_{i_{k}}-\bflambda_{i_{k}}),~\forall
i_{k}\in\mathcal{I}_{k}\label{OptimalSolutionV}
\end{align}
where
$\Hb^{k}=\left[\{\hb_{j_{l}}^{k}\}_{j_{l}\in\mathcal{I}}\right]\in\Cplx^{MQ_{k}\times|\mathcal{I}|}$,
$\Kb^{i_{k}}=\left[\{\Kb_{j_{l}}^{i_{k}}\}_{j_{l}\in\mathcal{I}}\right]^{T}\in\Cplx^{|\mathcal{I}|}$,
$\bfmu^{i_{k}}=[\{\mu_{j_{l}}^{i_{k}}\}_{j_{l}\in\mathcal{I}}]^{T}\in\Cplx^{|\mathcal{I}|}$,
and
$\bflambda_{i_{k}}=[(\bflambda_{i_{k}}^{1_{k}})^{T},\ldots,(\bflambda_{i_{k}}^{Q_{k}})^{T}]^{T}\in\Cplx^{MQ_{k}}$,
with $\bflambda_{i_{k}}^{q_{k}}\in\Cplx^{M}$ being the $i_{k}$-th block
of $\bflambda^{q_{k}}$. Hence, the optimization variable block $\vb$ can
be optimally solved in closed-form as well.

\subsection{Discussions}
\subsubsection{Computational Costs}
As noted above, each step of Algorithm1 can be carried out in
closed-form, which makes Algorithm 1 highly efficient. Specifically,
the most computational intensive operation in Algorithm 1 is the matrix inversion
\eqref{OptimalSolutionV}, which has complexity in the order of
O$((MQ_{k})^{3})$. However, this operation only needs to be computed
once for each cell $k$. As a result, compared to the standard
interior point algorithm, which has a per iteration complexity in the
order O$((\sum_{k\in\mathcal{K}}MQ_{k}I_{k})^{3})$, the proposed
ADMM approach has a cheaper per iteration computational cost, especially when
$|\mathcal{Q}|$ and $|\mathcal{I}|$ are large}.  %However, we should
%note that before the primal iterate satisfies the equality
%constraints in \eqref{NewVariables}, it can only serve as a high
%quality approximate solution for problem
%\eqref{PowerMinSingleStageRelax}.% And by Theorem
%\ref{ConvergenceADMM}, the obtained $\vb^\star$ (and {$\wb^\star$})
%would eventually converge to the optimal solution of problem
%\eqref{PowerMinSingleStageRelax} satisfying \eqref{NewVariables}.}

%Besides the advantage in saving the computational cost, the proposed
%algorithm can also be implemented in a distributive fashion. This is
%desirable since in practice, it is not practical to assume a central
%controller exists that solve problem \eqref{PowerMinSingleStage}.

%Suppose that no central
%controller is available that manages the entire network, but rather
%a central operator exists in each cell, which decides the BF for all
%the users in that cell. This can be observed by the fact that except

\subsubsection{Distributed Implementation}
Another advantage of the proposed algorithm is that it can be
implemented without a central controller. Observe that except
for $\{\Kb, \mbox{\boldmath$\kappa$}\}$, the computation for the
rest of the primal and dual variables can be performed  within each
cell without any information exchange among the cells. When updating
$\Kb$ and $\bfkappa$, each cell $k$ exchanges
$|\mathcal{I}_{k}||\mathcal{I}|$ complex values
$\{(\hb_{j_{l}}^{k})^{H}\vb_{i_{k}}|j_{l}\in\mathcal{I},i_{k}\in\mathcal{I}_{k}\}$
with the rest of cells. %This is possible
%since each cell operator $k$ can locally estimate the channel
%information $\{\hb_{j_{l}}^{k}|j_{l}\in\mathcal{I}\}$ by channel
%reciprocal property when TDD system is applied or channel estimation
%techniques in FDD system.
Once this is done, the subproblems \eqref{ProblemKSOCP} for updating
$\{\Kb, \mbox{\boldmath$\kappa$}\}$ can be again solved
independently by each cell. In conclusion, the ADMM approach allows
problem \eqref{PowerMinSingleStageRelax} to be solved in a distributed manner
across cells without a central operator.

\subsubsection{The Debiasing Step} After problem \eqref{PowerMinSingleStageRelax} is solved, performing
an additional ``de-biasing" step can further minimize the total power
consumption. That is, with the given set of selected active BSs
computed by the proposed single-stage ADMM approach, we can solve
problem \eqref{PowerMinSingleStageRelax} again, this time {\it
without} the sparse promoting terms. This can be done by making the
following changes to the proposed algorithm: {\it 1)} letting
$\beta_{q_{k}}=0,~~\forall  q_{k}\in\mathcal{Q}$; {\it 2)} setting
$\theta=1$; {\it 3)} only optimize over BSs with
$\vb^{q_{k}\star}\neq 0$. See reference \cite{Wright09} for further
justification of using such de-biasing technique in solving
regularized optimization problems.

\subsubsection{The Special Case of Power Minimization Problem}
As a byproduct of the proposed ADMM approach, the conventional power
minimization problem \eqref{PowerMin} without active BS selection
can also be efficiently solved using a simplified version of
Algorithm 1, by setting $\beta_{q_{k}}=0,~~\forall
q_{k}\in\mathcal{Q}$, and $\theta=1$. Compared to the existing
approaches for solving the same problem, the proposed ADMM approach is
computationally more efficient. For example, the uplink-downlink
duality approach \cite{Dahrouj10} needs to perform  matrix
inversion operations with complexity O$((MQ_{k})^{3})$ in each
iteration. The other ADMM based algorithms for solving problem
\eqref{PowerMin} either needs to solve
SDPs \cite{Shen12} or SOCPs \cite{Joshi12} in each iteration. %Thus,
%the proposed AMDD approach has computational benefit for problem
%\eqref{PowerMin} by its low complexity closed-form update rule
%compared to the existing methods when only power minimization is
%considered.
In contrast, by a novel splitting of the primal variables according to
the special structure of \eqref{PowerMin}, our proposed ADMM approach (i.e., Algorithm 1)
does not solve expensive subproblems; the subproblems are all solvable in closed forms.

%\subsection{Further Reduction of the Number of Activated
%BSs}\label{sec:reweight}

\subsubsection{Further Reduction of the Number of Active
BSs}\label{sec:reweight}

To achieve the maximum reduction of the number of active BSs, we
propose to adaptively reweight the coefficients
$\beta_{q_{k}},~~\forall  q_{k}\in\mathcal{Q}$. This reweighting
technique is popular in the compressive sensing literature to
increase the sparsity level of the solution; see e.g.,
\cite{Candes08,Mehanna12}. This can be done by first solving problem
\eqref{PowerMinSingleStageRelax}, and then updating the coefficient
$\beta_{q_{k}}$ by
\begin{align}
\beta_{q_{k}}&\longleftarrow\frac{\beta_{q_{k}}^{(0)}}{\|\wb^{q_{k}\star}\|+\epsilon},~~\forall
q_{k}\in\mathcal{Q},
\end{align}
where $\beta_{q_{k}}^{(0)}$, $\forall q_{k}\in\mathcal{Q}$, are the initial $\beta_{q_{k}}$ of problem \eqref{PowerMinSingleStageRelax} and $\epsilon>0$ is a small prescribed parameter to provide the
stability when $\|\wb^{q_{k}\star}\|$ is too small. With this new
set of $\beta_{q_{k}}$, \eqref{PowerMinSingleStageRelax} is solved
again. Intuitively, those BFs that have smaller magnitude will be
penalized more heavily in the comming iteration, thus is more likely
to be set to zero. In our numerical experiments to be shown in Sec.
\ref{sec:simuanddiss}, indeed we observe that by using such
reweighting technique, the number of active BSs converges very fast
and is much smaller than that obtained by solving problem
\eqref{PowerMinSingleStageRelax} only once.

%Before closing this section, we should also emphasize that since our
%goal in problem \eqref{PowerMinSingleStageRelax} is to approximately
%maximize the reduction of the number of activated BSs as in the
%original problem \eqref{PowerMinSingleStage}, only using $\ell_2$
%norm to approximate $\ell_0$ norm with fixed prescribed
%$\beta_{q_{k}},~~\forall  q_{k}\in\mathcal{Q}$, might not be enough.

\section{Sum Rate Maximization with Base Station Activation}\label{Sec:SumRate}
\subsection{Problem Formulation}\label{subsec:SumRateFormulation}

In this section, we show that how BS activation can be incorporated
into the design criteria {\bf C2)}, i.e., maximize the sum rate
subject to power constraint. We first note that, as explained in
Sec. \ref{sec:sysmodel}, even without considering BS activation,
solving sum rate maximization problem \eqref{SumRateMax} is itself strongly
NP-hard. Since this problem remains NP-hard regardless the
number of antennas at each user, we will consider a more general
scenario in which both BSs and users are equipped with multiple
antennas.

For simplicity of notation, we assume that all users have $N$
receive antennas. Let us change the notation of channel from
$\hb_{i_{k}}^{q_{l}}$ to $\Hb_{i_{k}}^{q_{l}}\in\Cplx^{N\times M}$.
In this way, the achievable rate for user $i_{k}$ becomes
\begin{align}
R_{i_{k}}({\bf v})=\log&\det\bigg({\bf I}+{\bf H}_{i_{k}}^{k}{\bf
v}_{i_{k}}{\bf v}_{i_{k}}^{H}({\bf
H}_{i_{k}}^{k})^{H}\big(\sum_{(l,j)\neq(k,i)}{\bf H}_{i_{k}}^{l}{\bf v}_{jl}{\bf
v}_{jl}^{H}({\bf H}_{i_{k}}^{l})^{H}+\sigma_{i_{k}}^{2}{\bf
I}\big)^{-1}\bigg).
\end{align}

Similar to the previous section, we aim at jointly maximizing
the sum rate and selecting the active BSs. To this end, we first
split the transmit BF ${\bf v}_{i_{k}}^{q_{k}}$ by ${\bf
v}_{i_{k}}^{q_{k}}=\alpha_{q_{k}}\bar{\bf v}_{i_{k}}^{q_{k}}$, with
$\alpha_{q_{k}}\in[0,1]$ representing whether BS $q_{k}$ is switched
on. That is, when $\alpha_{q_{k}}=0$, BS $q_k$ is switched off,
otherwise, BS $q_{k}$ is turned on. In the sequel, we will consider
the following single-stage regularized sum rate maximization problem
\begin{align}\label{SumRateMaxSplit}
\max_{\mbox{\boldmath $\alpha$},\bar{\bf v}}~&
\sum_{k\in\mathcal{K}}\sum_{i_{k}\in\mathcal{I}_{k}}R_{i_{k}}({\bf
v})-\sum_{q_{k}\in\mathcal{Q}}\mu_{q_{k}}\|\alpha_{q_{k}}\|_0\notag\\
{\rm s. t.}~&\alpha_{q_{k}}^2(\bar{\bf v}^{q_{k}})^{H}\bar{\bf
v}^{q_{k}}\leq P_{q_{k}},~~\forall  q_{k}\in\mathcal{Q}_{k},
\end{align}
where $\mu_{q_{k}}\geq 0,~~\forall  q_{k}\in\mathcal{Q}$, is the
parameter controlling the size of active BSs; $\mbox{\boldmath
$\alpha$}\triangleq\{\mbox{\boldmath$\alpha$}_{k}|k\in\mathcal{K}\}$
with $\mbox{\boldmath
$\alpha$}_{k}\triangleq[\alpha_{1_{k}},\alpha_{2_{k}},\ldots,\alpha_{Q_{k}}]^{T}\in\Ral^{Q_{k}}$.

Before further discussing how to deal with problem
\eqref{SumRateMaxSplit}, we will explain our motivation for
introducing the penalization term
$\sum_{q_{k}\in\mathcal{Q}}\mu_{q_{k}}\|\alpha_{q_{k}}\|_0$.
\begin{lemma}
{\it Let ($\bfalpha^{\star}$, $\bar\vb^{\star}$) denote the optimal
solution for \eqref{SumRateMaxSplit}. Then at optimality of problem
\eqref{SumRateMaxSplit}, each active BS $q_{k}$ contributes at least
$\mu_{q_{k}}$ bits/sec to the total achieved sum rate. Furthermore,
among all feasible solutions with the size of the active BS equals
to $\|\bfalpha^{\star}\|_0$, if $\mu_{q_{k}}=\mu$, $\forall q_{k}\in\mathcal{Q}$, ($\bfalpha^{\star}$, $\bar\vb^{\star}$)
gives the maximum sum rate.}
\end{lemma}
$\it Proof$ Define the optimal active BS set as
$Q^{\star}\triangleq\{\hat q_{ k}|\alpha_{\hat q_{k}}^{\star}>0\}$,
and denote the sum rate achieved at the optimal solution as
$R^{\star}$. Suppose BS $q_k$ is active at optimality, i.e.,
$q_{k}\in Q^{\star}$.  Let $\hat R^{\star}$ denotes the optimal
solution for problem \eqref{SumRateMaxSplit} with active BS set
$Q^{\star}-\{q_{k}\}$.

Suppose that $R^{\star}$ is no more than $\mu_{q_{k}}$ bits/sec
higher than $\hat R^{\star}$, i.e., $R^{\star}<\hat
R^{\star}+\mu_{q_{k}}$. This implies that
\begin{align*}
R^{\star}-\sum_{\hat q_{k}\in\mathcal{Q}}\mu_{\hat
q_{k}}\|\alpha_{\hat q_{k}}^{\star}\|_{0}< \hat
R^{\star}-\left(\sum_{\hat q_{k}\in\mathcal{Q}}\mu_{\hat
q_{k}}\|\alpha_{\hat q_{k}}^{\star}\|_{0}-\mu_{q_{k}}\right)
\end{align*}
which contradicts the optimality of the solution
($\bfalpha^{\star}$, $\bar\vb^{\star}$). The last claim is also easy
to see by a contradiction argument. \hfill \ensuremath{\Box}

%However, the $\ell_0$ norm in problem \eqref{SumRateMaxSplit} is
%non-convex and being hard to deal with. Thus, as in the previous
%section,

Unfortunately, the $\ell_0$ norm is not only non-convex but also not
continuous. As a result it is difficult to find even a locally
optimal solution for problem \eqref{SumRateMaxSplit}. Similar to
the previous section, we will relax, in the following, the $\ell_0$
norm to the $\ell_1$ norm. In the way, the regularized sum rate
maximization problem becomes
\begin{align}\label{SumRateMaxGroupLassoSplit}
\max_{\mbox{\boldmath $\alpha$},\bar{\bf v}}~&
\sum_{k\in\mathcal{K}}\sum_{i_{k}\in\mathcal{I}_{k}}R_{i_{k}}({\bf
v})-\sum_{q_{k}\in\mathcal{Q}}\mu_{q_{k}}|\alpha_{q_{k}}|\notag\\
{\rm s. t.}~&\alpha_{q_{k}}^2(\bar{\bf v}^{q_{k}})^{H}\bar{\bf
v}^{q_{k}}\leq P_{q_{k}},~~\forall  q_{k}\in\mathcal{Q}_{k},
\end{align}
In what follows, we will propose an efficient algorithm to compute a
stationary solution for this relaxed problem.

\begin{Rmk} Instead of splitting ${\bf v}_{i_{k}}^{q_{k}}$ and
penalizing $\|\mbox{\boldmath $\alpha$}_{k}\|_1$, another natural
modification is to add a group LASSO regularization term for each
BS's BF directly, i.e., use the regularization term $ \|{\bf
v}^{q_{k}}\|$ for BS $q_{k}$ in the objective function of problem
\eqref{SumRateMax}. However, when the power used by BS $q_{k}$ is
large, the magnitude of penalization term can dominate that of the
system sum rate. Thus solving such group-LASSO penalized problem
would effectively force the BSs to use only a small portion of its
power budget, which could lead to a dramatic reduction of the system
sum rate. The regularization in \eqref{SumRateMaxGroupLassoSplit}
avoids this problem.
\end{Rmk}

\subsection{Active BS Selection via a Sparse WMMSE Algorithm}\label{subsec:RedBS}
 By using a similar argument as in \cite[Proposition
1]{Hong12}, we can show that the penalized sum rate maximization
problem \eqref{SumRateMaxGroupLassoSplit} is equivalent to the
following penalized weighted mean square error (MSE) minimization
problem
\begin{subequations}
\begin{align}
\min_{\mbox{\boldmath $\alpha$},\bar{\bf v}, {\bf u},{\bf w}}&f({\bf
v},{\bf w},{\bf u})+\sum_{q_{k}\in\mathcal{Q}}\mu_{q_{k}}|\alpha_{q_{k}}|\\
{\rm s.t.}~&f({\bf v},{\bf w},{\bf
u})=\sum_{i_{k}\in\mathcal{I}}w_{i_{k}}e_{i_k}({\bf u}_{i_{k}},{\bf
v})-\log(w_{ik})\\
&\alpha_{q_{k}}^2(\bar{\bf v}^{q_{k}})^{H}\bar{\bf v}^{q_{k}}\leq
P_{q_{k}},~~\forall
q_{k}\in\mathcal{Q}_{k},~k\in\mathcal{K}.\label{eqPowerConstraint}
\end{align}
\end{subequations}
In the above expression, ${\bf u}\triangleq\{{\bf
u}_{i_{k}}\mid i_{k}\in\mathcal{I}\}$ is the set of all receive BFs of
the users; ${\bf w}\triangleq \{w_{i_{k}}|i_{k}\in\mathcal{I}\}$ is
the set of non-negative weights; $e_{i_k}$ is the MSE for estimating
$s_{i_{k}}$:
\begin{align}
\!\!\!\!e_{i_k}({\bf u}_{i_{k}},&{\bf v})\triangleq(1-{\bf u}^H_{i_k}{\bf H}^{k}_{i_k}{\bf v}_{i_k})(1-{\bf u}^H_{i_k}{\bf H}^{k}_{i_k}{\bf v}_{i_k})^{H}+\sum_{(\ell,j)\ne (k,i)}{\bf u}^H_{i_k}{\bf
H}^\ell_{i_k}{\bf v}_{j_\ell}{\bf v}_{j_\ell}^H({\bf
H}^\ell_{i_k})^H{\bf u}_{i_k}+\sigma^2_{i_k}{\bf u}^H_{i_k}{\bf
u}_{i_k}.
\end{align}

To guarantee convergence of the proposed algorithm, we further
replace the power constraint \eqref{eqPowerConstraint} by a slightly
more conservative constraint, namely $(\bar{\bf
v}^{q_{k}})^{H}\bar{\bf v}^{q_{k}}\leq P_{q_{k}},~
\alpha_{q_{k}}^{2}\leq 1$. The precise reason for doing so will be
explained shortly in the reasoning of Theorem \ref{Thm:Converge}. In
this way, the modified penalized weighted MSE minimization problem
for active BS selection is given by
\begin{center}
\fbox{\parbox[]{.76\linewidth}{ \noindent
\begin{align}\label{MainProblemSplit}
\min_{\mbox{\boldmath $\alpha$},\bar{\bf v}, {\bf u},{\bf w}}&f({\bf
v},{\bf w},{\bf u})+\sum_{q_{k}\in\mathcal{Q}}\mu_{q_{k}}|\alpha_{q_{k}}|\\
{\rm s.t.}&~f({\bf v},{\bf w},{\bf
u})=\sum_{i_{k}\in\mathcal{I}}w_{i_{k}}e_{i_k}({\bf u}_{i_{k}},{\bf
v})-\log(w_{ik})\nonumber\\
&(\bar{\bf
v}^{q_{k}})^{H}\bar{\bf v}^{q_{k}}\leq P_{q_{k}},\nonumber\\
&\alpha_{q_{k}}^{2}\leq 1,~~\forall  q_{k}\in\mathcal{Q}_{k}.\notag
\end{align}
 }}
\end{center}
Although the modified power constraint will shrink the original
feasible set whenever $\alpha_{q_{k}}^2\neq0$ or $\pm 1$, thus may
reduce the sum rate performance of the obtained transceiver, our
numerical experiments (to be shown in Section \ref{sec:simuanddiss})
suggest that satisfactory sum rate performance can still be
achieved.

Due to the fact that problem \eqref{MainProblemSplit} is convex in
each block variables, global minimum can be obtained for each block
variable when fixing the rest. Furthermore, the problem is strongly
convex for block ${\bf u}$ and ${\bf w}$, respectively, and the
unique optimal solution ${\bf u}_{i_{k}}^{\star}$ and
$w_{i_{k}}^{\star}$, $~\forall  i_{k}\in\mathcal{I}$, can be obtained
in closed-form:
\begin{align}
{\bf u}_{i_{k}}^{\star}({\bf v})&=\left(\sum_{j_{l}\in\mathcal{I}}{\bf
H}_{i_{k}}^{l}{\bf v}_{j_{l}}{\bf v}_{j_{l}}^{H}({\bf
H}_{i_{k}}^{l})^{H}+\sigma_{i_{k}}^{2}{\bf I}\right)^{-1}{\bf
H}_{i_{k}}^{k}{\bf v}_{i_{k}},\notag\\
&\triangleq {\bf J}_{i_{k}}^{-1}({\bf v}){\bf H}_{i_{k}}^{k}{\bf
v}_{i_{k}}\label{UMMSE}\\
w_{i_{k}}^{\star}({\bf v})&=\left(1-{\bf v}_{i_{k}}^{H}\left({\bf
H}_{i_{k}}^{k}\right)^{H}{\bf J}_{i_{k}}^{-1}({\bf v}){\bf
H}_{i_{k}}^{k}{\bf v}_{i_{k}}\right)^{-1}.\label{WMMSE}
\end{align}
On the other hand, problem \eqref{MainProblemSplit} can also be
rewritten as
\begin{align}\label{MainProblemSplitIndicator}
\min_{\mbox{\boldmath $\alpha$},\bar{\bf v}, {\bf u},{\bf w}}&f({\bf
v},{\bf w},{\bf
u})+\sum_{q_{k}\in\mathcal{Q}}\mu_{q_{k}}|\alpha_{q_{k}}|+I_{1}(\bar{\bf
v})+I_{2}(\mbox{\boldmath$\alpha$})
\end{align}
where $I_{1}(\bar{\bf v})$ and $I_{2}(\mbox{\boldmath $\alpha$})$
are indicator functions for both constraints defined respectively as
\begin{align*}
I_{1}(\bar{\bf v})&=\left\{\begin{array}{ll} 0,
~&\mbox{if}~(\bar{\bf
v}^{q_{k}})^{H}\bar{\bf v}^{q_{k}}\leq P_{q_{k}},~~\forall  q_{k}\in\mathcal{Q}_{k},\\
\infty,~&\mbox{otherwise}
\end{array}\right.,\\
I_{2}(\mbox{\boldmath $\alpha$})&=\left\{\begin{array}{ll}
0,~&\mbox{if}~\alpha_{q_{k}}^{2}\leq 1,~~\forall  q_{k}\in\mathcal{Q}_{k},\\
\infty,~&\mbox{otherwise}
\end{array}\right..
\end{align*}
Observe that when the problem is written in the form of
\eqref{MainProblemSplitIndicator}, all its nonsmooth parts are {\it
separable} across block variables $\mbox{\boldmath $\alpha$}$,
$\bar{\bf v}$, $\bf u$, and $\bf w$. Such separability is guaranteed
by our modified power constraints, and is referred to as the
``regularity condition" for nonsmooth optimization; see
\cite{Tseng01} for details about this condition. Combining this
property with the fact that at most two blocks, namely
$\mbox{\boldmath $\alpha$}$ and $\bar{\bf v}$, may not have unique
minimizer, a block coordinate descent (BCD) procedure \footnote{In
our context, the BCD procedure refers to the computation strategy
that cyclically updates the blocks ${\bf u}$, ${\bf w}$,
$\bar{\bf v}$, and $\mbox{\boldmath $\alpha$}$ one at a time.} is
guaranteed to converge to the stationary point of problem
\eqref{MainProblemSplit}. This is proven by Lemma 3.1 and Theorem
4.1 of \cite{Tseng01}. The following theorem summarizes the
preceding discussion.
\begin{Theorem}\label{Thm:Converge}
A BCD procedure that iteratively optimizes problem
\eqref{MainProblemSplit} for each block variables $\bf u$, $\bf w$,
$\bar{\bf v}$, and $\mbox{\boldmath $\alpha$}$, can always converge
to a stationary solution of problem \eqref{MainProblemSplit}.
\end{Theorem}

In the following, we discuss in detail how problem
\eqref{MainProblemSplit} can be solved for each block variables in
an efficient manner. For blocks ${\bf u}$ and ${\bf w}$, optimal
solutions are shown in \eqref{UMMSE} and \eqref{WMMSE},
respectively. For the optimization problem of $\mbox{\boldmath
$\alpha$}$, notice that when fixing $({\bf u}, {\bf w}, {\bf
\bar{v}})$, the objective of problem \eqref{MainProblemSplit} is
separable among the cells. Therefore $K$ independent subproblems can
be solved simultaneously, with the $k$-th subproblem assuming the
following form
\begin{align}\label{AlphaUpdate}
\min_{\mbox{\boldmath $\alpha$}_{k}}~&(\mbox{\boldmath
$\alpha$}_{k})^{T}{\bf A}_{k}\mbox{\boldmath $\alpha$}_{k}-2Re({\bf
b}_{k}^{H}\mbox{\boldmath $\alpha$}_{k})+\sum_{q_{k}\in\mathcal{Q}}\mu_{q_{k}}|\alpha_{q_{k}}|\notag\\
{\rm s. t.}~&\alpha_{q_{k}}^{2}\leq 1,~~\forall
q_{k}\in\mathcal{Q}_{k}
\end{align}
where
\begin{align*}
{\bf A}_{k}&\triangleq \sum_{i_{k}\in\mathcal{I}_{k}}{\rm
diag}(\bar{\bf v}_{i_{k}})^{H}\left(\sum_{j_{l}\in\mathcal{I}}w_{j_{l}}({\bf
H}_{j_{l}}^{k})^{H}{\bf u}_{j_{l}}{\bf u}_{j_{l}}^{H}{\bf
H}_{j_{l}}^{k}\right){\rm diag}(\bar{\bf v}_{i_{k}})\\
{\bf b}_{k}&\triangleq \sum_{i_{k}\in\mathcal{I}_{k}}w_{i_{k}}{\rm
diag}(\bar{\bf v}_{i_{k}})^{H}({\bf H}_{i_{k}}^{k})^{H}{\bf
u}_{i_{k}}.
\end{align*}

Problem \eqref{AlphaUpdate} is a quadratically constrained LASSO
problem.  It can be solved optimally by again applying a BCD
procedure, with the block variables given by $\alpha_{q_{k}}$,
$~\forall  q_{k}\in\mathcal{Q}_{k}$ (e.g., \cite{Wright09}). For the
$q_k$-th block, its optimal solution $\alpha_{q_{k}}^{\star}$ must
satisfy the following first-order optimality condition
\begin{align}
&2(c_{q_{k}}-({\bf
A}_{k}[q,q]+\gamma_{q_{k}}^{\star})\alpha_{q_{k}}^{\star}) \in
\mu_{q_{k}}\partial|\alpha_{q_{k}}^{\star}|,\label{Subgradient}\\
&\gamma_{q_{k}}^{\star}\geq 0,~(1-(\alpha_{q_{k}}^{\star})^{2})\ge
0\label{Constraints}\\
&(1-(\alpha_{q_{k}}^{\star})^{2})\gamma_{q_{k}}^{\star}=0\label{Complementarity}
\end{align}
where $\gamma_{q_{k}}^{\star}$ is the optimal dual variable for the
$q_{k}$th power constraint of problem \eqref{AlphaUpdate}, and
$c_{q_{k}}\triangleq \mbox{Re}({\bf b}_{k}[q])-\sum_{p\neq q}{\bf
A}_{k}[p,q]\alpha_{p_{k}}$. Therefore, when
$2\left|c_{q_{k}}\right|\leq \mu_{q_{k}}$, we have
$\alpha_{q_{k}}^{\star}=0$ . In the following, let us focus on the
case where $2|c_{q_{k}}|>\mu_{q_{k}}$. In this case, from the
expression of the subgradient \eqref{Subgradient}, we have
$\alpha_{q_{k}}^{\star}=\frac{-\mu_{q_{k}}{\rm
sign}(\alpha_{q_{k}}^{\star})+2c_{q_{k}}}{2({\bf
A}_{k}[q,q]+\gamma_{q_{k}}^{\star})}$. Since
$\gamma_{q_{k}}^{\star}\geq 0$, ${\bf A}_{k}[q,q]\geq0$, and
$2|c_{q_{k}}|>\mu_{q_{k}}$, we have ${\rm
sign}(\alpha_{q_{k}}^{\star})={\rm sign}(c_{q_{k}})$. By plugging
$\alpha_{q_{k}}^{\star}$ into the objective function of problem
\eqref{AlphaUpdate}, it can be shown the objective value is an
increasing function of $\gamma_{q_{k}}^{\star}$. Therefore, by the
monotonicity of $\gamma_{q_{k}}^{\star}$, primal and dual
constraints \eqref{Constraints}, and the complementarity condition
\eqref{Complementarity}, in the case of $2|c_{q_{k}}|>\mu_{q_{k}}$,
$\alpha_{q_{k}}^{\star}$ has the following structure
\begin{align}\label{alphaupdate}
\alpha_{q_{k}}^{\star}=\left\{\begin{array}{ll}
\frac{-\mu_{q_{k}}{\rm sign}(c_{q_{k}})+2c_{q_{k}}}{2{\bf
A}_{k}[q,q]}, & \mbox{if}\left|\frac{-\mu_{q_{k}}{\rm
sign}(c_{q_{k}})+2c_{q_{k}}}{2{\bf
A}_{k}[q,q]}\right|<1\\
{\rm sign}(c_{q_{k}}), &\mbox{otherwise}
\end{array}\right.
\end{align}

Similarly, when fixing $(\mbox{\boldmath $\alpha$}, {\bf w}, {\bf
u})$, the optimization problem for $\vb$ is convex and separable
among $K$ cells, and the $k$-th subproblem is expressed as
\begin{align}\label{VUpdate}
\min_{\bar\vb_{i_{k}},~i_{k}\in\mathcal{I}_{k}}~&\sum_{i_{k}\in\mathcal{I}_{k}}\left(\bar\vb_{i_{k}}^{H}{\bf
C}_{k}\bar\vb_{i_{k}}-\bar\vb_{i_{k}}^{H}{\bf D}_{i_{k}}-{\bf
D}_{i_{k}}^{H}\bar\vb_{i_{k}}\right)\notag\\
{\rm s. t.}~& \sum_{i_{k}\in\mathcal{I}_{k}}
(\bar\vb_{i_{k}}^{q_{k}})^{H}\bar\vb_{i_{k}}^{q_{k}}\leq
P_{q_{k}},~~\forall  q_{k}\in\mathcal{Q}_{k},
\end{align}
where \begin{align*} &{\bf C}_{k}\triangleq \mbox{\boldmath
$\hat\alpha$}_{k}\left(\sum_{j_{l}\in\mathcal{I}}w_{j_{l}}({\bf
H}_{j_{l}}^{k})^{H}{\bf u}_{j_{l}}{\bf u}_{j_{l}}^{H}{\bf
H}_{j_{l}}^{k}\right)\mbox{\boldmath
$\hat\alpha$}_{k}\in\Cplx^{Q_{k}M\times Q_{k}M},\\
&{\bf D}_{i_{k}}\triangleq
w_{i_{k}}\mbox{\boldmath
$\hat\alpha$}_{k}({\bf H}_{i_{k}}^{k})^{H}{\bf
u}_{i_{k}}\in\Cplx^{Q_{k}M},~~\forall
i_{k}\in\mathcal{I}_{k},\\
&\mbox{\boldmath $\hat\alpha$}_{k}\triangleq{\rm diag}({\bf
\alpha}_{1_{k}}{\bf I},\ldots,{\bf \alpha}_{Q_{k}}{\bf
I})\in\Cplx^{Q_{k}M\times Q_{k}M}.
\end{align*}

We wish to efficiently solve the problem by iteratively
updating its block components $\bar{\bf v}^{q_{k}}$, $~\forall
q_{k}\in\mathcal{Q}_{k}$. However, as discussed in
Theorem \ref{Thm:Converge}, the algorithm convergence requires that the
optimization problem has at most two block components which do not have unique optimal solution. To
furfill this requirement, we add a regularization term
$\sum_{q_{k}\in\mathcal{Q}_{k}}\epsilon
(\bar\vb^{q_{k}})^{H}\bar\vb^{q_{k}}$ to the objection function of
problem \eqref{VUpdate} with $\epsilon>0$. Thus, when
$\epsilon\rightarrow 0$, the solution for the BF $\bar{\bf
v}^{q_{k}\star}$ can be obtained by checking the first order
optimality condition, and this can be expressed as
\begin{align}\label{VUpdateRule}
\bar{\bf v}_{i_{k}}^{q_{k}\star}(\delta_{q_{k}})=&\left( {\bf
C}_{k}[q_{k},q_{k}]+\delta_{q_{k}}^{\star}{\bf
I}\right)^{\dag}\bigg({\bf D}_{i_{k}}[q_{k}]-\sum_{j_{k}\neq
q_{k}}{\bf C}_{k}[q_{k},j_{k}]\bar{\bf
v}_{i_{k}}^{j_{k}\star}\bigg),~~\forall  i_{k}\in\mathcal{I}_{k}.
\end{align}
In the above expression, $\dag$ denotes the Moore-Penrose
pseudoinverse; $\delta_{q_{k}}^{\star}\geq 0$ is the optimal dual
variable for the $q_k$-th power constraint;  ${\bf
C}_{k}[q_{k},j_{k}]\in\Cplx^{M\times M}$ and ${\bf
D}_{i_{k}}[q_{k}]\in\Cplx^{M}$ are, respectively, subblocks of
matrices ${\bf C}_{k}$ and ${\bf D}_{i_{k}}$. By the complementarity
condition, $\delta_{q_{k}}^{\star}=0$ if $(\bar{\bf
v}^{q_{k}\star}(0))^{H}\bar{\bf v}^{q_{k}\star}(0)\leq P_{q_{k}}$.
Otherwise, it should satisfy $(\bar{\bf
v}^{q_{k}\star}(\delta_{q_{k}}^{\star}))^{H}\bar{\bf
v}^{q_{k}\star}(\delta_{q_{k}}^{\star})= P_{q_{k}}$. For the latter
case, $\delta_{q_{k}}^{\star}$ can be found by a simple bisection
method.

In summary, our main algorithm can be summarized in the following
table.
\begin{center}
\fbox{\parbox[]{.99\linewidth}{ \noindent {\bf Sparse WMMSE (S-WMMSE)
algorithm}:
\begin{algorithmic}[1]
\State {\bf Initialization} Generate a feasible set of variables
$\{\bar{\bf v}_{i_{k}}\},~i_{k}\in\mathcal{I}$, and let
$\alpha_{q_{k}}=1~~\forall  q_{k}\in\mathcal{Q}_{k},k\in\mathcal{K}$.

\State {\bf Repeat}

\State ~~~${\bf u}_{i_{k}}\leftarrow {\bf J}_{i_{k}}^{-1}({\bf
v}){\bf H}_{i_{k}}^{k}{\bf v}_{i_{k}},~~\forall  i_{k}\in\mathcal{I}$

\State ~~~$w_{i_{k}}\leftarrow (1-{\bf v}_{i_{k}}^{H}\left({\bf
H}_{i_{k}}^{k}\right)^{H}{\bf J}_{i_{k}}^{-1}({\bf v}){\bf
H}_{i_{k}}^{k}{\bf v}_{i_{k}})^{-1},~~\forall  i_{k}\in\mathcal{I}$

\State ~~~$\bar{\bf v}^{q_{k}} $ is iteratively updated by
 \eqref{VUpdateRule}, $~\forall
q_{k}\in\mathcal{Q}_{k}$, $~\forall  k\in\mathcal{K}$

\State ~~~$\alpha_{q_{k}}$ is iteratively updated by
$$\alpha_{q_{k}}=\left\{\begin{array}{ll}
0,&\mbox{if}~2|c_{q_{k}}|\leq\mu_{q_{k}}\\
\eqref{alphaupdate},&\mbox{otherwise}
\end{array}\right.,~~\forall
q_{k}\in\mathcal{Q}_{k},~k\in\mathcal{K}$$

\State{\bf Until} Desired stopping criteria is met

\end{algorithmic}
}}
\end{center}

Similar to what we have done in the previous section, the de-biasing
and reweighting procedures can further improve the sum rate
performance and decrease the number of active BSs, respectively. The
de-biasing procedure utilizes the given set of active BSs computed by
the S-WMMSE algorithm, and solve problem \eqref{MainProblemSplit}
again, this time {\it without} the sparse promoting terms. In
particular we make the following changes to the S-WMMSE algorithm:
{\it 1)} letting $\mu_{q_{k}}=0$ for each $q_{k}\in\mathcal{Q}$;
{\it 2)} skipping step 6; {\it 3)} setting $\alpha_{q_{k}}={\rm
sign}(\alpha_{q_{k}}^{\star})$, $~\forall  q_{k}$. In the reweighting
procedure, we iteratively apply S-WMMSE to the reweighted problem
with the parameter $\mu_{q_{k}}$ being updated by
\begin{align}
\mu_{q_{k}}\longleftarrow\frac{\mu_{q_{k}}^{(0)}}{|\alpha_{q_{k}}|+\epsilon},~~\forall
q_{k}\in\mathcal{Q},
\end{align}
where $\mu_{q_{k}}^{(0)}$, $\forall q_{k}\in\mathcal{Q}$, are the initial $\mu_{q_{k}}$ of problem \eqref{MainProblemSplit}.

Furthermore, the proposed S-WMMSE algorithm can be solved
distributively among each cell, under the following assumptions: i)
there is a central controller in each cell; ii) the central
controller for cell $k$ has the CSI $\Hb_{j_{l}}^{k},~\forall
j_{l}\in\mathcal{I}$ and iii) each user $i_{k}\in\mathcal{I}$
can locally estimate the received signal plus noise covariance
matrix ${\bf J}_{i_{k}}$ and the received channel matrix
$\Hb_{i_{k}}^{k}$. The last assumption ensures that user $i_{k}$ can
update $\ub_{i_{k}}$ and $w_{i_{k}}$ locally. After updating
$\ub_{i_{k}}$ and $w_{i_{k}}$, each user $i_k$ can broadcast them to all the
central controllers. Combined with assumption ii), the central
controller in cell $k$ can then update $\bar v^{q_{k}}$ and
$\alpha_{q_{k}}$, $~\forall  q_{k}\in\mathcal{Q}_{k}$.

\subsection{Joint active BS selection and BS clustering}\label{subsec:RedBSAndCluster}
In addition to controlling the number of active BSs, we can further
optimize the size of BS clusters by adding an additional
penalization on the BFs. Specifically, since ${\bf
v}_{i_{k}}^{q_{k}}$ being zero means user $i_{k}$ is not served by
BS $q_{k}$, it follows that user $i_{k}$ is served with a small BS
cluster means $\|{\bf v}_{i_{k}}^{q_{k}}\|$ is nonzero for only a
few $q_{k}$s. Thus, a set of group LASSO regularization terms,
$\sum_{q_{k}\in\mathcal{Q}_{k}}\left\|{\bf
v}_{i_{k}}^{q_{k}}\right\|$,$i_k\in\mathcal{I}$, can be added to the
objective function of problem \eqref{SumRateMax} to reduce the size
of BS clusters; see \cite{Hong12} for details. Hence, to jointly
control the size of BS cluster and reducing the BS usage, the
objective function of the penalized weighted MMSE minimization
problem \eqref{MainProblemSplit} is now modified as
\begin{align}\label{MainProblemSplitModified}
f({\bf v},{\bf w},{\bf
u})+\sum_{k\in\mathcal{K}}\left(\sum_{i_{k}\in\mathcal{I}_{k}}\lambda_{k}\sum_{q_{k}\in\mathcal{Q}_{k}}\|\bar{\bf
v}_{i_{k}}^{q_{k}}\|\right)+\sum_{q_{k}\in\mathcal{Q}}\mu_{q_{k}}|\alpha_{q_{k}}|,
\end{align}
where $\lambda_{k}\geq 0, ~\forall  k\in\mathcal{K}$, is the parameter
to control the size of BS cluster in cell $k$. For this modified
problem, again a BCD procedure with block variables,
$\mbox{\boldmath $\alpha$}$, $\bar{\bf v}$, ${\bf u}$, and ${\bf
w}$, can be used to compute a locally optimal solution. The only
difference from the algorithm proposed in the previous section is
the computation of $\bar{\bf v}$. This can be carried out by solving
a quadratically constrained group LASSO problem. See in \cite[Table
I ]{Hong12} for details.

\section{Simulation Results}\label{sec:simuanddiss}
%%%%%%%%%%%%%%%%%%%%%%%%%%%%%%%%%%%%%%%%%%%%%%%%%%%%%%%%%%%%%%%%%%%%%%%
%%%%%%%%%%%%%%%%%%%%%%%%%%%%%%%%%%%%%%%%%%%%%%%%%%%%%%%%%%%%%%%%%%%%%%%
\begin{figure}[t] \centering
     \resizebox{0.6\textwidth}{!}{\includegraphics{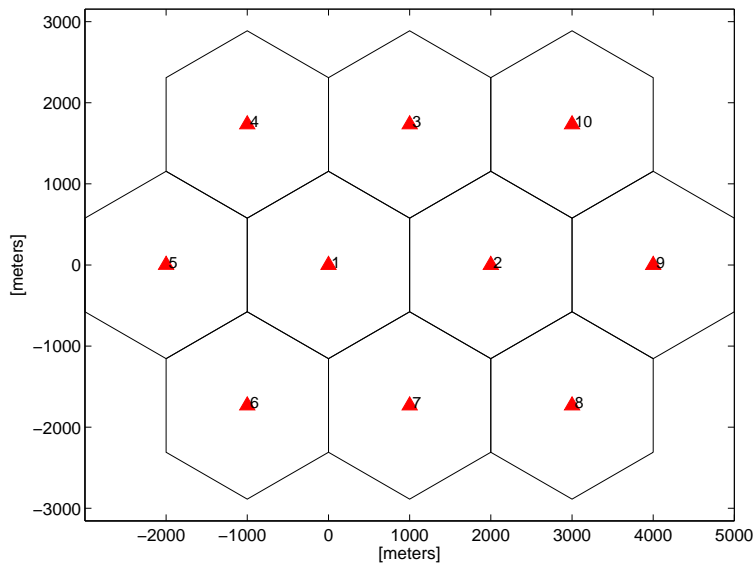}}
      \caption{\footnotesize Network configuration}
     \label{fig:fig1}

\end{figure}
%%%%%%%%%%%%%%%%%%%%%%%%%%%%%%%%%%%%%%%%%%%%%%%%%%%%%%%%%%%%%%%%%%%%%%%
%%%%%%%%%%%%%%%%%%%%%%%%%%%%%%%%%%%%%%%%%%%%%%%%%%%%%%%%%%%%%%%%%%%%%%%

%%%%%%%%%%%%%%%%%%%%%%%%%%%%%%%%%%%%%%%%%%%%%%%%%%%%%%%%%%%%%%%%%%%%%%%
%%%%%%%%%%%%%%%%%%%%%%%%%%%%%%%%%%%%%%%%%%%%%%%%%%%%%%%%%%%%%%%%%%%%%%%
\begin{figure}[t] \centering
     \resizebox{0.6\textwidth}{!}{\includegraphics{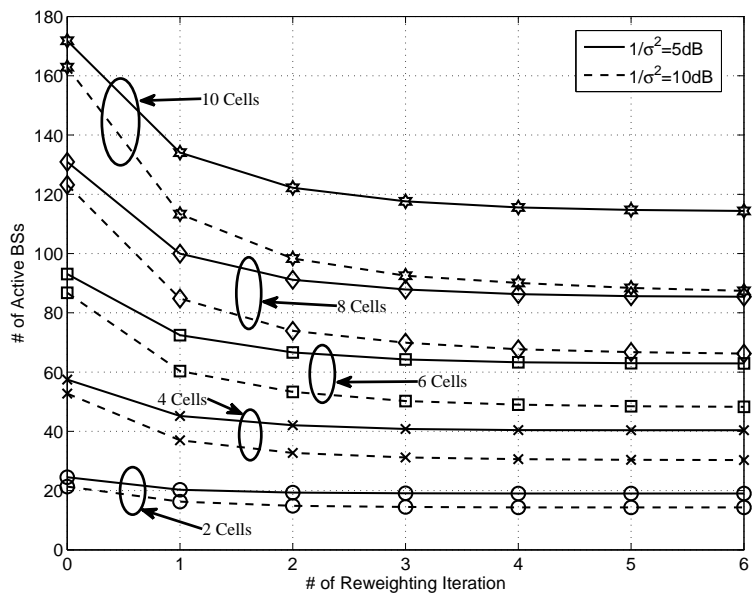}}
      \caption{\footnotesize Number of active BSs after each reweighting procedure.}
     \label{fig:fig2}

\end{figure}
%%%%%%%%%%%%%%%%%%%%%%%%%%%%%%%%%%%%%%%%%%%%%%%%%%%%%%%%%%%%%%%%%%%%%%%
%%%%%%%%%%%%%%%%%%%%%%%%%%%%%%%%%%%%%%%%%%%%%%%%%%%%%%%%%%%%%%%%%%%%%%%

%%%%%%%%%%%%%%%%%%%%%%%%%%%%%%%%%%%%%%%%%%%%%%%%%%%%%%%%%%%%%%%%%%%%%%%
%%%%%%%%%%%%%%%%%%%%%%%%%%%%%%%%%%%%%%%%%%%%%%%%%%%%%%%%%%%%%%%%%%%%%%%
\begin{figure}[t] \centering
     \resizebox{0.6\textwidth}{!}{\includegraphics{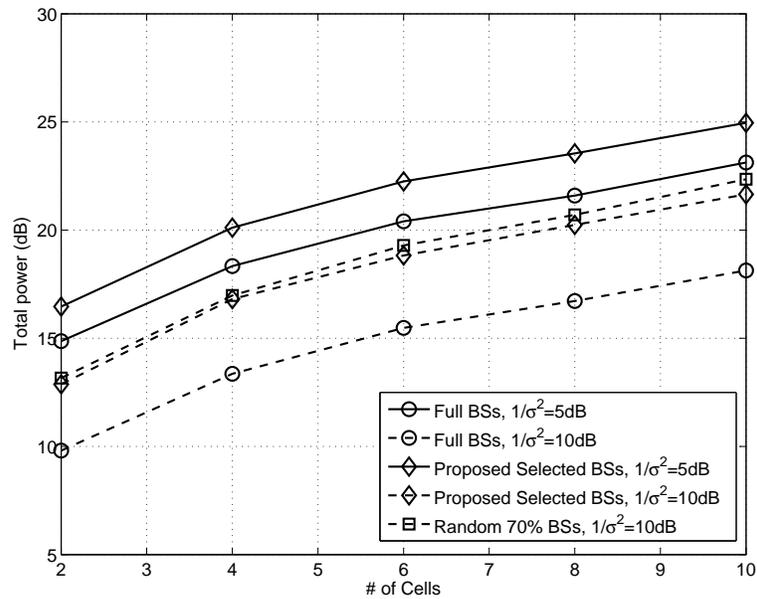}}
      \caption{\footnotesize Power consumption for all scenarios considered.}
     \label{fig:fig3}

\end{figure}
%%%%%%%%%%%%%%%%%%%%%%%%%%%%%%%%%%%%%%%%%%%%%%%%%%%%%%%%%%%%%%%%%%%%%%%
%%%%%%%%%%%%%%%%%%%%%%%%%%%%%%%%%%%%%%%%%%%%%%%%%%%%%%%%%%%%%%%%%%%%%%%

%%%%%%%%%%%%%%%%%%%%%%%%%%%%%%%%%%%%%%%%%%%%%%%%%%%%%%%%%%%%%%%%%%%%%%%
%%%%%%%%%%%%%%%%%%%%%%%%%%%%%%%%%%%%%%%%%%%%%%%%%%%%%%%%%%%%%%%%%%%%%%%
\begin{figure}[t] \centering
     \resizebox{0.6\textwidth}{!}{\includegraphics{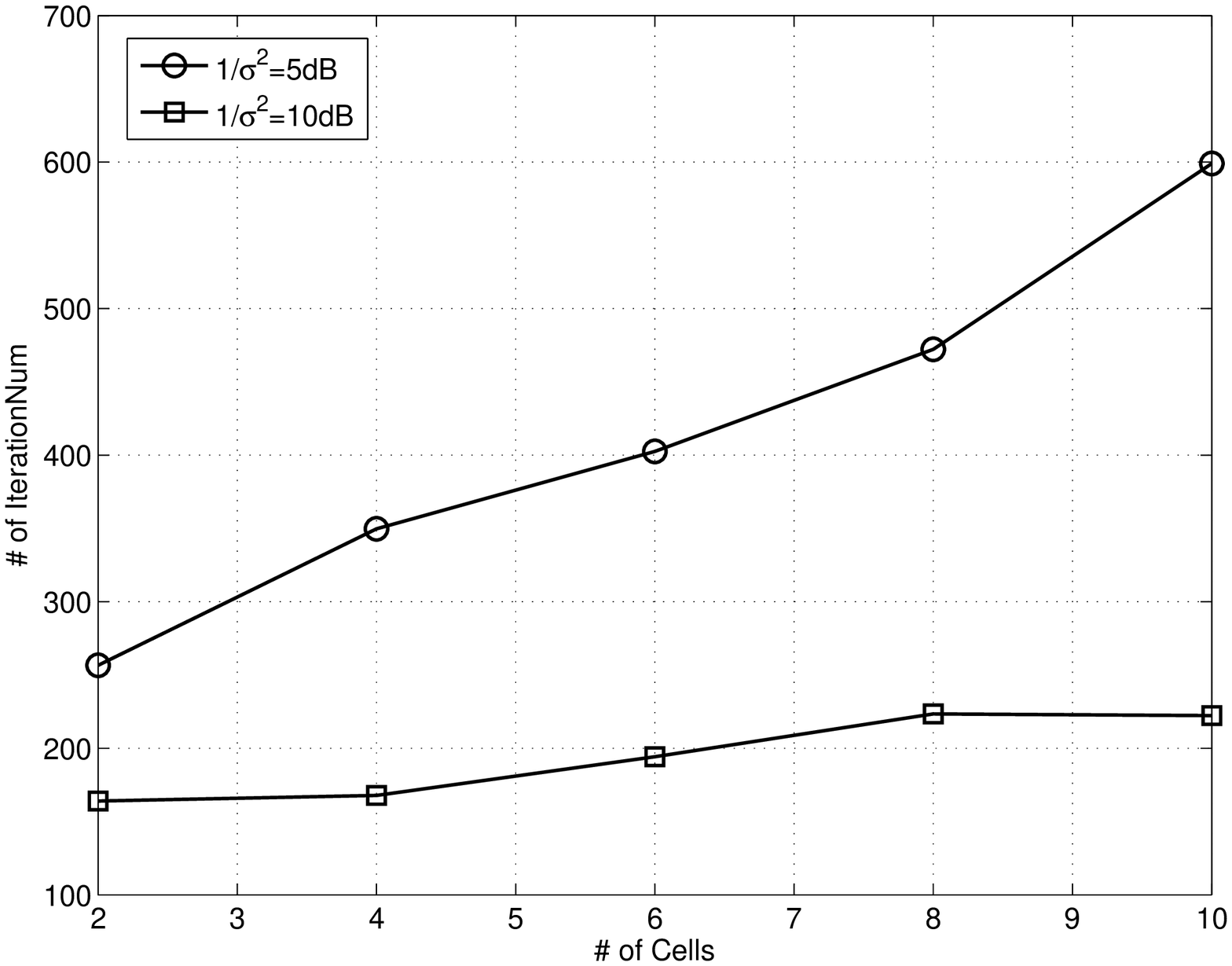}}
      \caption{\footnotesize The required number of ADMM iterations for the scenario where all the BSs are active.}
     \label{fig:fig4}

\end{figure}
%%%%%%%%%%%%%%%%%%%%%%%%%%%%%%%%%%%%%%%%%%%%%%%%%%%%%%%%%%%%%%%%%%%%%%%
%%%%%%%%%%%%%%%%%%%%%%%%%%%%%%%%%%%%%%%%%%%%%%%%%%%%%%%%%%%%%%%%%%%%%%%

In the following numerical experiments, we consider HetNets with at
most $10$ cells. The distance between centers of adjacent cells is
set as $2000$ meters; see Fig. \ref{fig:fig1} for an illustration of
the network configuration. In each cell, we place one BS at the center of the
cell (representing the macro BS), and randomly and uniformly
place $I$ users and $Q-1$ remaining BSs. The channel model we use is
Rayleigh channel with zero mean and variance
$(200/d_{i_{k}}^{q_{l}})^3 L_{i_{k}}^{q_{l}},$ where
$d_{i_{k}}^{q_{l}}$ is the distance between BS $q_{l}$ and user
$i_{k}$, and $10\log10(L_{i_{k}}^{q_{l}})\sim N(0,64)$. We also
assume that $\sigma_{i_{k}}^{2}=\sigma^{2},~~\forall
i_{k}\in\mathcal{I}$. All the simulation results are averaged over
$100$ channel realizations. The results shown for problem
\eqref{PowerMinSingleStageRelax}, \eqref{MainProblemSplit} and
\eqref{MainProblemSplitModified} are those obtained after performing
the de-biasing step. The proposed algorithm is compared to the
following two scenarios: 1) all the BSs are turned on; 2) in each
cell, the central BS and a randomly selected fixed number of the
remaining BSs are turned on. Note that for both of these cases, full
JP is used within each cell. Clearly, the first scenario can serve
as the performance upper bound, and the latter can serve as a
reasonable heuristic algorithm to select active BSs since BSs and
users are uniformly distributed in each cell.

In the first set of simulations, the total power minimization design
criterion is considered. We set $I=10$, $Q=20$, $M=5$, and
$\tau_{i_{k}}=15$dB, $~\forall  i_{k}\in\mathcal{I}$. Furthermore, we
assume that the power budget for BSs in the center of each cell is
$10$ dB while the budget for the rest of the BSs is set to be $5$dB.
We apply the ADMM approach to solve the proposed formulation
\eqref{PowerMinSingleStageRelax} with reweighting procedure. Since
the objective QoS $\tau_{i_{k}},~~\forall  i_{k}\in\mathcal{I}$ may
not always be feasible, we declare that this realization is infeasible if a particular problem realization cannot converge within $2000$
ADMM iterations. We select the stepsize as $\rho=5$, and use the
following stopping criterion
\begin{align*}
&\max\left(\left\|\frac{\|vec(\Kb)\|_{\infty}}{\max(1,\|\Kb\|_{F})}\right\|,
\left\|\frac{\vb-\wb}{\max(1,\|\vb\|,\|\wb\|)}\right\|_{\infty},\max_{i_{k}\in\mathcal{I}}(|\kappa_{i_{k}}^{2}-\sigma^{2}|),\frac{f^{\min}(\wb^{(t)})-f^{\min}(\wb^{(t-1)})}{f^{\min}(\wb^{(t-1)})}\right)<10^{-4}.
\end{align*}
In Fig. \ref{fig:fig2}, we plot the number of active BSs after each
reweighting procedure on $\beta_{q_{k}},~~\forall
q_{k}\in\mathcal{Q}$ for $1/\sigma^{n}=5$dB and $10$dB,
respectively. From this figure, it can be observed that the number
of active BSs decreases fast for the first $2$ reweighting
iterations, and converges within $6$ reweighting iterations. In Fig.
\ref{fig:fig3}, the obtained minimum total power is plotted against
the number of cells. We can observe that the minimum required power
for BSs selected by the proposed formulation
\eqref{PowerMinSingleStageRelax} is more than that achieved by
activating all the BSs in each cell. This is reasonable since the latter serves as a lower
bound of achievable power consumption.  On the other hand, when
$1/\sigma^{2}=10$dB, we compare the minimum power consumption
achieved by the following two networks: i) a network with $70\%$ of
randomly activated BSs (the center BSs in each cell are always
active) and ii) the network optimized by the proposed algorithm
($35.8\%\sim43.45\%$ of BSs are activated for each number of cells).
It can be observed that the proposed formulation is able to use
much smaller number of BSs with similar total transmit power to
support the same set of QoS constraints. This demonstrates the
efficacy of the proposed method. Additionally, Fig. \ref{fig:fig4}
plots the required number of ADMM iterations for the power
minimization only design \eqref{PowerMin} (with all BSs being turned
on). We observe that the proposed ADMM approach converges fairly
fast. Note that the convergence speed depends on the channel
quality, $\sigma^{2}$: when the channel condition is good enough,
i.e., $1/\sigma^{2}=10$dB, it converges within $250$ ADMM
iterations.

%%%%%%%%%%%%%%%%%%%%%%%%%%%%%%%%%%%%%%%%%%%%%%%%%%%%%%%%%%%%%%%%%%%%%%%
%%%%%%%%%%%%%%%%%%%%%%%%%%%%%%%%%%%%%%%%%%%%%%%%%%%%%%%%%%%%%%%%%%%%%%%
\begin{figure}[t] \centering
     \resizebox{0.6\textwidth}{!}{\includegraphics{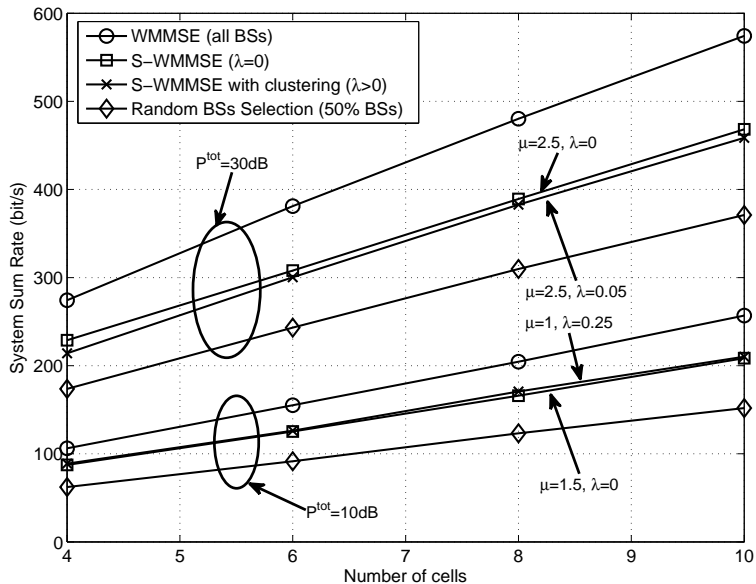}}
      \caption{\footnotesize The comparison on sum rate performance over different number of cells and total power budgets, $P^{tot}$, between proposed S-WMMSE algorithm, the performance upper bound, and a heuristic random selection.}
     \label{fig:fig5}

\end{figure}
%%%%%%%%%%%%%%%%%%%%%%%%%%%%%%%%%%%%%%%%%%%%%%%%%%%%%%%%%%%%%%%%%%%%%%%
%%%%%%%%%%%%%%%%%%%%%%%%%%%%%%%%%%%%%%%%%%%%%%%%%%%%%%%%%%%%%%%%%%%%%%%

\begin{table}\centering
\begin{tabular}{|l|c|c|c|c|}
\hline Number of Cells & 4 & 6 & 8 & 10 \\
\hline

WMMSE (all BSs) & 40 & 60 & 80 & 100 \\
\hline
Random BSs Selection ($50\%$ BSs) & 20 & 30&  40&  50\\
\hline
S-WMMSE ($\mu=1.5$, $\lambda=0$), $P^{tot}=10$dB & 18.27 & 26.33 & 35.24 & 43.53 \\
\hline
S-WMMSE ($\mu=1$, $\lambda=0.25$), $P^{tot}=10$dB & 20.18 & 28.67 & 38.51 & 47.04 \\
\hline
S-WMMSE ($\mu=2.5$, $\lambda=0$), $P^{tot}=30$dB & 21.11 & 28.38 & 36.42 & 45.80 \\
\hline
S-WMMSE ($\mu=2.5$, $\lambda=0.05$), $P^{tot}=30$dB & 20.21 & 28.73 & 37.80 & 46.95 \\
\hline
\end{tabular}
\caption{The number of active BSs v.s.\
different number of cells.}\label{tab:tab1}
\end{table}

%%%%%%%%%%%%%%%%%%%%%%%%%%%%%%%%%%%%%%%%%%%%%%%%%%%%%%%%%%%%%%%%%%%%%%%%
%%%%%%%%%%%%%%%%%%%%%%%%%%%%%%%%%%%%%%%%%%%%%%%%%%%%%%%%%%%%%%%%%%%%%%%%
\begin{figure}[t]
\begin{center}
{\subfigure[][$P^{tot}=10$dB]{\resizebox{.45\textwidth}{!}{\includegraphics{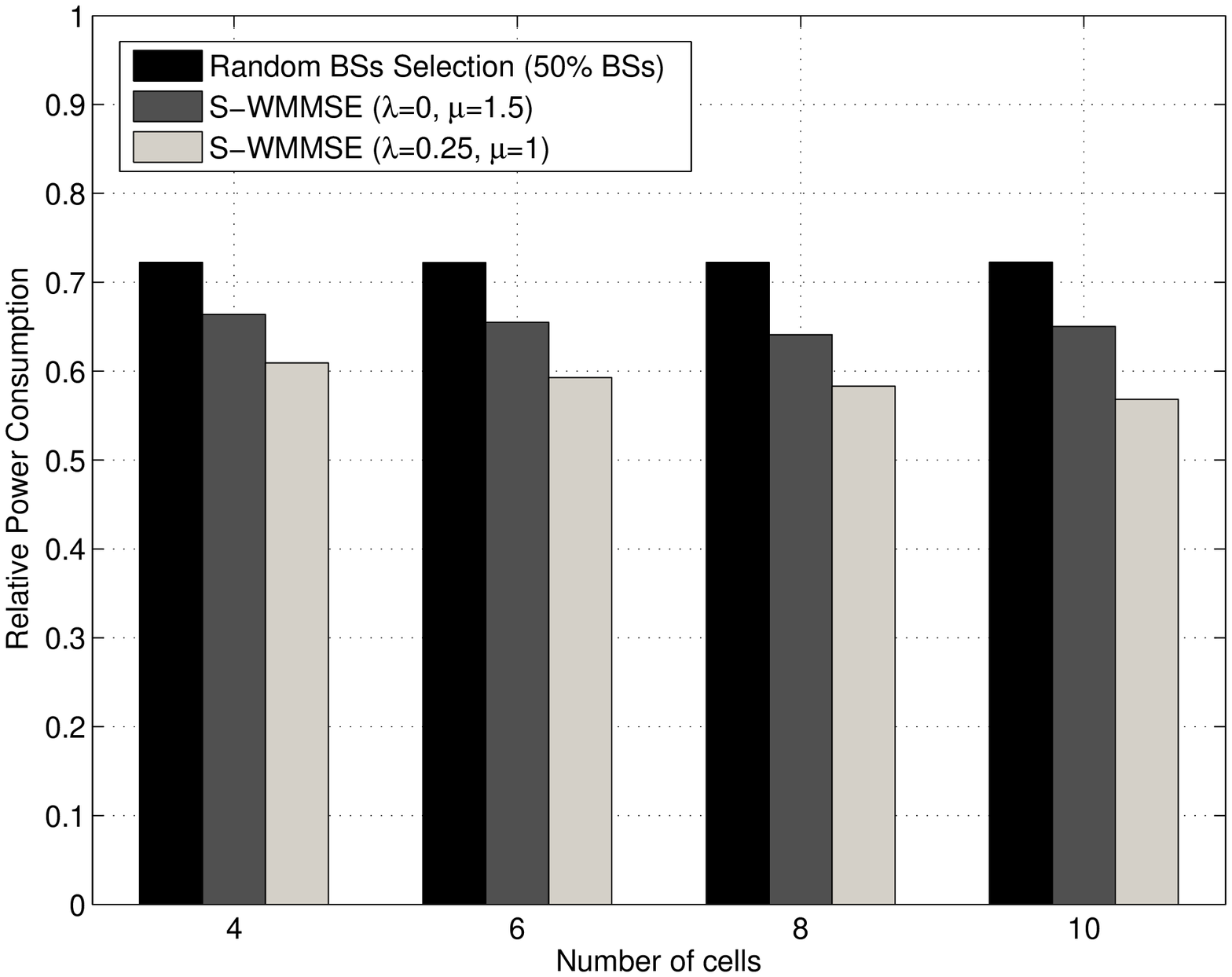}}}}
\hspace{1pc}
{\subfigure[][$P^{tot}=30$dB]{\resizebox{.45\textwidth}{!}{\includegraphics{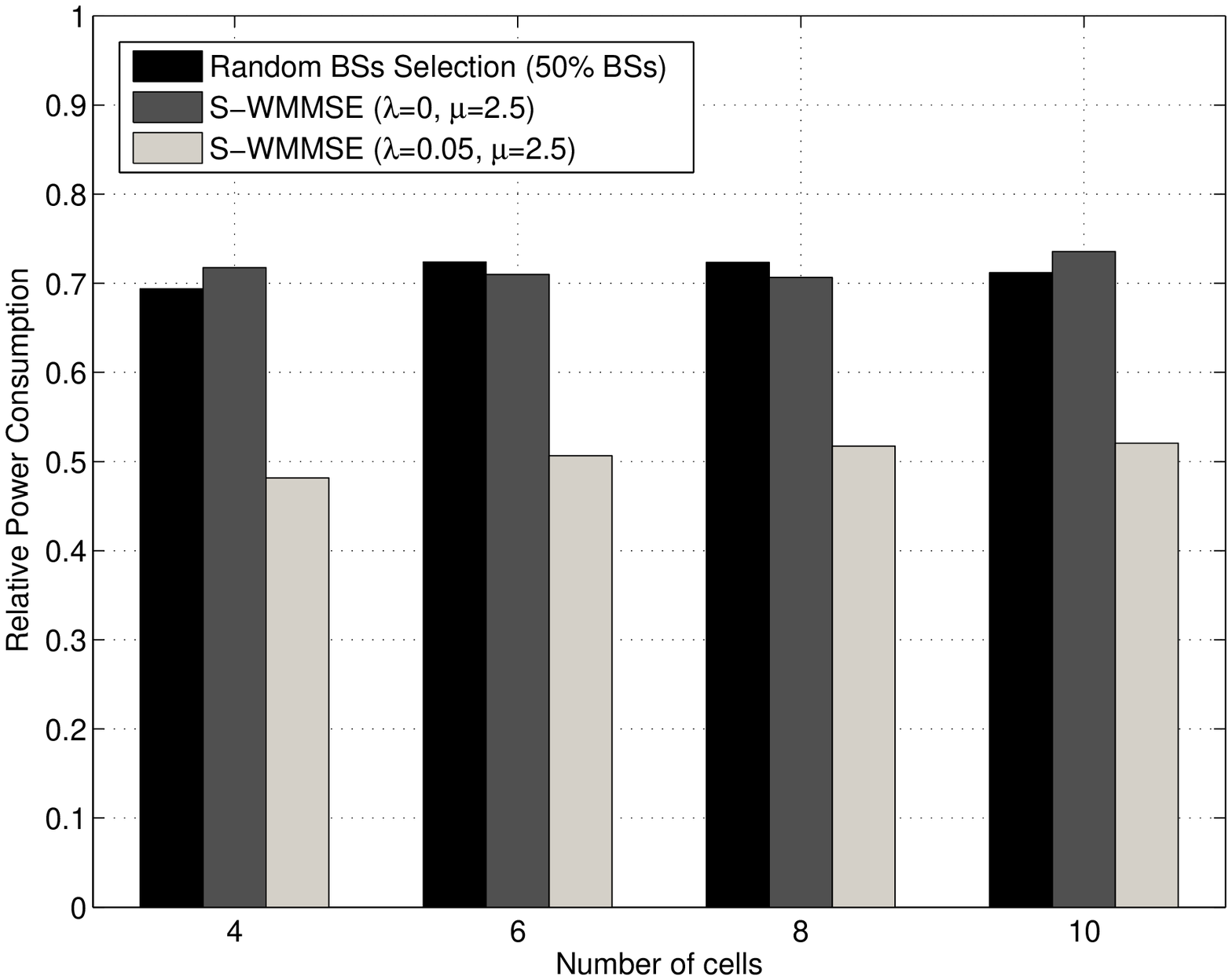}}}}
\end{center}
\caption{Comparison of the power consumption for different schemes
with varying $P^{tot}$. The total power used for the case where all
BSs are active is normalized to 1.}\label{fig:fig6}
\end{figure}

In the second simulation set, the sum rate maximization design
criterion is investigated. Let $I=10$, $Q=10$, $M=4$, $N=2$ and
$P^{tot}$ denote the total power budget in each cell. The power
budget for BSs located in the center of the cells is $P^{tot}/2$,
and the rest of the BSs have equal power budgets. For simplicity, we
set $\mu_{q_{k}}=\mu,~~\forall  q_{k}\in\mathcal{Q}$,
$\lambda_{k}=\lambda,~~\forall  k\in\mathcal{K}$, and
$\sigma_{i_{k}}^{2}=1,~~\forall  i_{k}\in\mathcal{I}$. The reweighting
procedure is performed until no BS reduction is possible or less than
$50\%$ of BSs is active. This is for fair comparison with random
selection scheme turning on $50\%$ of BSs. In Fig. \ref{fig:fig5},
the system sum rate performance for the proposed S-WMMSE algorithm
is compared with $P^{tot}=10$dB and $30$dB. We can observe that
S-WMMSE can achieve about 80\% of the sum rate compared to the upper
bound while activating around 50\% BSs (see Tab. \ref{tab:tab1} for
details about the number of active BSs). Furthermore, while the
number of active BSs for S-WMMSE is about the same as the random
selection scheme, the S-WMMSE can still achieve more than 34\% and
23\% improvement in sum rate performance for $P^{tot}=10$dB and
$30$dB, respectively. It is worth noting that when BS clustering is
considered, there is no sizable decrease in the sum
rate performances. However, the total power consumption is
significantly reduced; see Fig. \ref{fig:fig6}. This is because when
optimizing the BS clustering, the coverage of each BS is reduced, so
does the interference level. As a result, less total transmit power
is able to support similar sum rate performance.

In summary, our simulation results suggest that for the power
minimization design criterion, the proposed ADMM approach can
effectively reduce the BS usage while minimizing the required
minimum power consumption. On the other hand, when considering the
sum rate maximization design criterion, the proposed S-WMMSE
algorithm can effectively reduce the BS usage and the size of BS
cluster simultaneously.

\section{Concluding Remarks}\label{sec:conclusion}

In this paper, we have considered the downlink beamforming problems that
jointly select the active BSs while {\bf C1)} minimizing the total power
consumption; or {\bf C2)} maximizing the sum rate performance. Since the
considered problems are shown to be strongly NP-hard in general, we have
utilized the sparse-promoting techniques and proposed formulations that
effectively select the active BSs. Moreover, for these two design
criteria, we have developed efficient distributed algorithms that
are based on respectively the ADMM algorithm and WMMSE algorithm.
Interestingly, when specialized to the standard problem of
minimum power MISO downlink beamforming without BS selection(see \cite{Wiesel06,Dahrouj10}),
our proposed ADMM approach is more efficient than the
conventional approach that exploits the uplink-downlink duality \cite{Rashid-Farrokhi98,Wiesel06,Dahrouj10} with computation complexity analysis.
For future work, it would be interesting to apply the ADMM approach to
efficiently solve general large-scale SOCPs, and to consider downlink beamforming problems and algorithms for
situations where only long-term channel statistics are available.

\appendix
\subsection{Proof of Theorem \ref{NPHard}}\label{subsec:NPHardProof}

To prove Theorem \ref{NPHard}, it is sufficient to show that problem
\eqref{PowerMinBSSelection} is strongly NP-hard. Consider a simple single-cell
network with $Q$ single antenna BSs serving $Q$ users. That is,
$K=1$, $M=1$, $|\cQ_k|=|\cI_k|=Q$. Then problem
\eqref{PowerMinBSSelection} can be simplified to
\begin{align}\label{NPHardProb}
\min_{\{p_{i}^{q}\}}~&\sum_{i=1}^{Q}\left\|\sum_{q=1}^{Q}p_{i}^{q}\right\|_{0}\notag\\
\rm{s. t.}~&\frac{\sum_{q=1}^{Q}p_{i}^{q}g_{i}^{q}}{\sigma_{i}^{2}+\sum_{j\neq i}\sum_{q=1}^{Q}p_{j}^{q}g_{i}^{q}}\geq \tau_{i},\\
& \sum_{i=1}^{Q}p_{i}^{q}\leq P_{q},~p_{i}^{q}\geq 0,~~\forall  ~i,q=1,\ldots,Q,\notag
\end{align}
where we have omitted the cell index $k$, and have
defined $p_{i}^{q}\triangleq\|\vb_{i}^{q}\|_2^2 \mbox{~and~}
g_{i}^{q}\triangleq\|\hb_{i}^{q}\|_2^2,~~\forall  i,q=1,\ldots,Q$. We
prove that problem \eqref{NPHardProb} is strongly NP-hard by
establishing a polynomial time transformation from the so-called
vertex cover problem. The vertex cover problem can be described as
follows: given a graph $\mathcal{G} = (\mathcal{V},\mathcal{E})$ and
a positive integer $N \leq |\mathcal{V}|$, we are asked whether
there exists a vertex cover of size $N$ or less, i.e., a subset
$\mathcal{V}'\subset\mathcal{V}$ such that $|\mathcal{V}'|\leq N$,
and for each edge $\{u,v\}\in\mathcal{E}$ at least one of $u$ and
$v$ belongs to $\mathcal{V}$.

Given a graph $\mathcal{G}=(\mathcal{V},\mathcal{E})$ with $|\mathcal{V}|=Q$, we let
\begin{align*}
&g_{i}^{q}=g_{q}^{i}=\left\{\begin{array}{ll}
1,~&\mbox{if}~i=q~\mbox{or}~(i,q)\in\mathcal{E}\\
0,~&\mbox{if}~(i,q)\not\in\mathcal{E}
\end{array}\right.\\
&\tau_{i}=\frac{1}{Q^{2}},~\sigma_{i}^{2}=Q,~P_{q}=Q,~~\forall  q=1,\ldots,Q.
\end{align*}
We claim that the optimal value of problem \eqref{NPHardProb} is less than or equal to
$N$ if and only if there exists a vertex cover set $\mathcal{V}'$ for the graph satisfying $|\mathcal{V}'|\leq N$.

\noindent
\emph{``If" direction:} Let $\mathcal{V}'$ with $|\mathcal{V}'|\leq N$ be the vertex cover set for the graph $\mathcal{G}$. Without loss of generality, suppose $\mathcal{V}'=\{1,2,\ldots,N\}$. Then we can construct a feasible solution for problem \eqref{NPHardProb} based on the cover set $\mathcal{V}'$ such that the optimal value of problem \eqref{NPHardProb} at this point is equal to $N$. In particular, we have
\begin{align*}
&p_{i}^{q}=1,~i=1,\ldots,Q,~q=1,2,\ldots,N\\
&p_{i}^{q}=0,~i=1,\ldots,Q,~q=N+1,N+2,\ldots,Q
\end{align*}
Next, we check the feasibility of the above constructed solution.
\begin{itemize}
\item For user $i=1,2,\ldots,N$, the SINR constraint in \eqref{NPHardProb} is satisfied, since $p_{q}^{q}=g_{q}^{q}=1$ for all $q=1,\ldots,N$.
\item For user $i=N+1,N+2,\ldots,Q$, according to the definition of the cover set, there must exist $q\in\mathcal{V}'$ such that $(i,q)\in\mathcal{E}$ and thus $p_{i}^{q}=g_{i}^{q}=1$. Hence, the SINR constraint of user $i=N+1,N+2,\ldots,Q$ are also satisfied.
\end{itemize}

\noindent
 {\it ``Only if" direction:} Suppose that the optimal value of problem \eqref{NPHardProb} is less than or equal to $N$ and its optimal solution is $p_{i}^{q\star},~~\forall  i,q=1,\ldots,Q$. We construct the following sets
\begin{align*}
S_{q}\triangleq\{i\mid p_{i}^{q\star}g_{i}^{q}>0\}=\{i\mid p_{i}^{q\star}>0\},~q=1,\ldots,Q.
\end{align*}
By the fact that the optimal value of problem \eqref{NPHardProb} is less than or equal to $N$, we know that at most $N$ of the defined sets $S_{q}$ are nonempty sets. Without loss of generality, suppose these $N$ nonempty sets are $S_{1},\ldots,S_{N}$. Furthermore, the fact that all SINR constraints are satisfied implies
\begin{align*}
\mathcal{V}=\bigcup_{q=1}^{Q}S_{q}=\bigcup_{q=1}^{N}S_{q}.
\end{align*}
The above shows that $\{1,2,\ldots,N\}$ constitutes a cover set of $\mathcal{V}$, which completes the proof. \hfill \ensuremath{\Box}

\bibliographystyle{IEEEbib}
%\footnotesize
\bibliography{refs13}

\end{document}